\documentclass[aps,preprint,nofootinbib,showpacs,showkeys,10pt]{revtex4-2}
\usepackage[utf8]{inputenc}
\usepackage[english]{babel}
\usepackage{amsmath,amsthm}            
\usepackage{extarrows}
\usepackage{amssymb}
\usepackage{nicefrac}
\usepackage{fancyref}
\usepackage{hyperref}
\usepackage{siunitx} 
\usepackage{graphicx}                   
\usepackage{subfigure}
\usepackage{float} 
\restylefloat{figure}
\restylefloat{table}
\usepackage{xcolor}
\usepackage{braket} 

\usepackage[compat=1.1.0]{tikz-feynman}

\newcommand{\slsh}[1]{\not \! #1}

\newcommand{\be}{\begin{equation}}
\newcommand{\ee}{\end{equation}}
\newcommand{\bea}{\begin{eqnarray}}
\newcommand{\eea}{\end{eqnarray}}

\begin{document}

\title{Interaction field strength between a scalar particle and two massless vector bosons \\ in presence of an external magnetic field}

\author{Jorge Jaber-Urquiza}

\email{jorgejaber@ciencias.unam.mx}

\author{Angel Sanchez}

\email{ansac@ciencias.unam.mx}

\affiliation{Facultad de Ciencias, Universidad Nacional Aut\' onoma de M\' exico, Apartado Postal 50-542, Ciudad de M\'exico 04510, M\'exico.}


\begin{abstract}

\noindent In this work we study the interaction strength among a neutral scalar boson and two massless vector bosons in presence of an external magnetic field. Based on global symmetries, we build the general tensor structure amplitude $\mathcal{M}^{\mu\nu}$, for the process $V^\mu+V^\nu\longrightarrow\phi$, in terms of the vector bosons polarization states. Then, we present a novel methodology to compute the one-loop amplitude contributions for an homogeneous magnetic field with arbitrary strength. With the obtained results, expressed in terms of integrals over Schwinger parameters, we explore its behavior in two regions, widely used in the literature, the strong and weak field strength regions. The methodology presented in this work can be employed to compute an arbitrary process in presence of an external magnetic field where the initial and final states are neutral.

\end{abstract}

\maketitle

\section{Introduction}
\label{sec.intro}

\noindent In several branches of physics, scalar fields are an important study object since they drive interesting physical phenomena at different energy scales: superconductivity in condensed matter~\cite{GinzLand}, color superconductivity and superfluidity in compact astrophysical objects~\cite{AstroObje1,AstroObje2,AstroObje3}, accelerated expansion at the early stages of the Universe~\cite{LINDE,MUKHA} or as mass-giver within the particle Standard Model~\cite{PrediccionHiggs1,PrediccionHiggs2}.

In recent decades, great theoretical effort and progress has been done in order to understand, with better precision, the properties of fundamental scalar fields. A remarkable example is the Higgs boson, where the studies of its different channels of decay ($H\longrightarrow\gamma\gamma$)~\cite{Ellis1,Kotsky,HiggsGamaGama,Marciano2012a} and production ($gg\longrightarrow H$)~\cite{Georgi,Spira,HiggsReport}, through massless vector bosons, led to its final discovery~\cite{ATLASdetect,CMSdetect,Confirma1,Confirma2,Confirma3}. On the other hand, for non-fundamental scalar fields, the decay into two photons has shown to be a relevant process in quantum chromodynamics (QCD) since it is used to probe the mesons' structure and represent an important contribution to the hadronic light-by-light scattering. The decaying of a neutral pion into two photons ($\pi^0\longrightarrow\gamma\gamma$), for example, test the flavor-singlet chiral symmetry~\cite{Pion1,Pion2}. In this way, the interaction among a scalar particle and two massless vector bosons is relevant for fundamental and composite fields.

Since the above mentioned processes, for scalar particles, can be realized in relativistic heavy ion collisions~\cite{Berguer,Enterria,PionHIC}, where the physical phenomena become enriched by the extreme conditions created, then, the interaction among the particles is expected to be modified. One condition recently accounted for, in peripheral collisions, is a magnetic field~\cite{Skokov,Zhong2014} whose presence drives new effects on the system~\cite{Kharzeev,Miransky2015}. At this point, a question that naturally emerges is: how does the external magnetic field change the scalar field interaction strength? This question has been addressed in Ref.~\cite{HiggsDecayMagnetic} where the effect of a magnetic background in the Higgs decay width into two photons is studied, considering correction coming from the electroweak sector of the Standard Model, and finding that the decay width has singularity when considering large magnetic field values.

In general, the presence of an external magnetic field effect has been studied by several authors, in a wider number of particle processes and physical observables, facing with quite involved calculations independently of the formalism used to introduce the magnetic field effect in the process: Schwinger's proper time or Ritus' eigenfunctions~\cite{SCWING,RITUS,FeInSa}.

In both formalisms it is possible to obtain expressions in terms of Landau levels which are useful to study physical situations where the highest energy scale is the magnetic field. In this latter scenario, called strong magnetic field limit, the calculations are enormously simplified since the only contributions come from the lowest Landau level (LLL) and a dimensional reduction improves the ultraviolet (UV) behavior~\cite{Gusynin}. However, in this limit, there is not a clear treatment for the UV-divergences, in some cases they seem to depend on the magnetic field~\cite{AyalaLuisDivergPionNeutral} contrasting with the results obtained with a different formalism, where they are magnetic field independent for an arbitrate magnetic field strength~\cite{SCWING,Scoccola2022b}. In an arbitrary field strength, the calculations in terms of Landau levels, are cumbersome~\cite{RefereeInd1,RefereeInd2,IndConductivity,IndLeptonProd,Shovkovy2022}, in part due to all levels contributing to the process~\cite{TaiwanDebil}.

\newpage

The Schwinger's proper time is used to explore both the strong~\cite{Gusynin} and weak~\cite{TaiwanDebil} magnetic field strength regions. The latter one is studied by performing a power series expansion in B (with B the magnetic field) keeping only the leading terms~\cite{AyalaDebilGluon,AyalaDebilPionMass}. In Refs.~\cite{Piccinelli,Jaber}, that study particles decays in the weak magnetic field region, it was shown that the momentum of the progenitor particle plays an important rôle in the processes~\cite{TSAI.1,TSAI.2}.

Although some analytical advance and physical insight is gained in the strong and weak magnetic field strength regions, other escapes from the analysis, for example, the rôle played by the Schwinger's phase or the different kinematic conditions. Schwinger's phase is specially relevant in process that involves charged asymptotic states~\cite{Scoccola2022a} or in quantum corrections to 3-point  functions~\cite{Skobelevv,TaiwanDebil,AyalaTriang2}.

An additional complication, introduced by the magnetic field presence, arises in the tensor structure decomposition in vector bosons correlation functions, as can be seen in the calculations of the photon~\cite{RITUS,Batalin,Hugo} or gluon~\cite{Hattori} polarization tensors, the photon splitting amplitude~\cite{Adler,RitusPapan,RitusPapan2}, the photon production through gluon fusion amplitude~\cite{AyalaTriangUltimo}, etc.

By using the Schwinger proper time operator approach \cite{Dittrich,SCWING}, in Ref.~\cite{HiggsDecayMagnetic} the decay rate of Higgs bosons to two photons in presence of an external magnetic field has been computed. The main part of that work focuses on the appearance of instabilities coming from the gauge sector of the electroweak of the Standard Model. Even though, the contribution coming from the quark is briefly discussed in Sec.~V, there is not shown any analytical expression for an arbitrary kinetical configuration for the photons. Moreover, the results presented seems to be computed without taking any regularization scheme and restrict the analysis to the strong field region for a particular photon momenta configuration.

In general, the calculations that involve charged quantum fluctuations, dressed by the magnetic field, are usually carried out by taking the sum over spins in the loop at early stages of the calculation, resulting in  long and involved expressions that make difficult any analytical treatment for an arbitrary field strength, as can be seen in the previous mentioned references, as well as in~\cite{Scoccola2019,Shovkovy2020,Shovkovy2021}. With the goal to obtain compact and analytical expressions for an arbitrary magnetic field strength, from which the strong and weak field regions in different kinematic regimes can be studied, this work presents a novel methodology that pretends to make more transparent and simple on how to deal with the complications presented in the calculations with a magnetic field as: the Schwinger's phase, the loop-momenta integration, the UV regularization and the spinorial trace treatment.

With context in mind, this work is organized as follows: in Sec.~\ref{sec.model}, we present a quantum electrodynamics (QED)-like model in which a scalar particle interact with two massless vector bosons through quantum corrections; in Sec.~\ref{sec.tensor}, using global symmetries, we build the tensor structure amplitude, for the scalar particle production through vector boson fusion process, within a magnetic field background. By computing the amplitude's one-loop contributions, in Sec.~\ref{sec.calculations} we present a methodology to incorporate an homogeneous magnetic field with arbitrary strength. With the exact expression obtained, in Sec.~ \ref{sec.aproximations} we study the amplitude's behavior in the strong and weak magnetic field strength regions and discuss the physical importance of the massless vector bosons kinematics. In Sec. \ref{sec.results}, we present results for the magnetic field effect on the scalar boson production through VB fusion within the weak field at low perpendicular momentum approximation. Finally, Sec.~\ref{sec.concl} contains our conclusions.

\section{Model}
\label{sec.model}

\noindent In order to explore the magnetic field influence on the scalar boson production (decay) through the fusion (emission) of two \textbf{massless vector bosons} (VB), let us start by considering a QED-based model, given by
\begin{equation}
	\mathcal{L}=-\frac{1}{4}B^{\mu\nu}B_{\mu\nu}+\bar{\psi}\left(i\slsh{D}-m\right)\hat{\psi}+\dfrac{1}{2}\partial^\mu\phi\partial_\mu\phi-\dfrac{1}{2}m_\phi^2\phi^2-h\phi\bar{\psi}\psi,
	\label{eq.lagrangian}
\end{equation}
where  $\psi(x)$ and $\phi(x)$ are a fermion and a scalar fields, and 
\begin{equation}
	\label{eq.covder}
	D^\mu\equiv\partial^\mu+igV^{\mu}(x)\ \ \mbox{and\  \ } 
	B^{\mu\nu}=\partial^\mu V^{\nu}(x)-\partial^\nu V^{\mu}(x),
\end{equation}
are the covariant derivative and the $V$-field strength tensor, respectively, with $V^\mu(x)$ an abelian gauge field.

\begin{figure}[H]
	\centering
	\includegraphics[width=5cm]{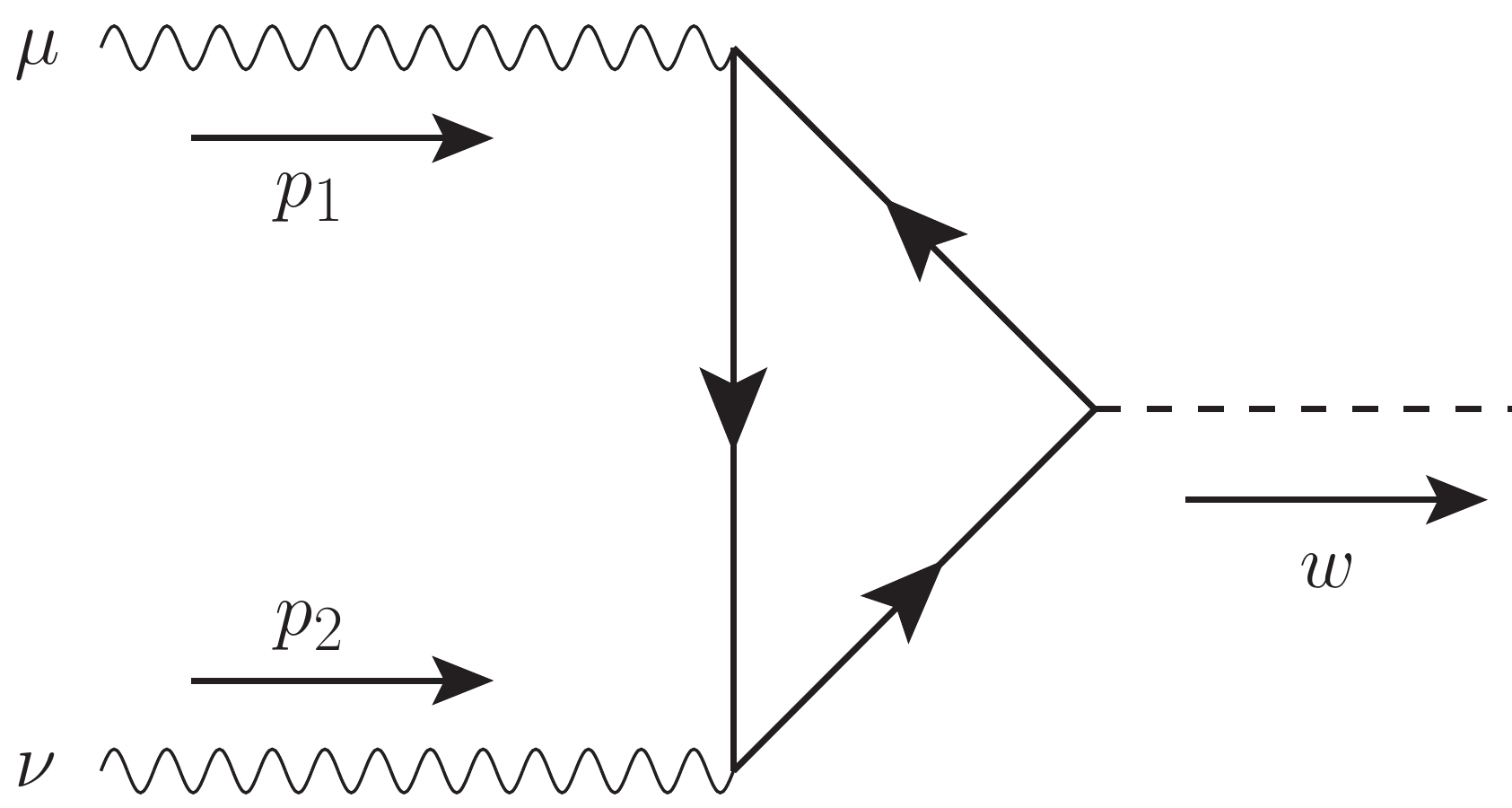}
	\caption{One-loop Feynman diagram that accounts the interaction between the vector and scalar sectors.}
	\label{figuratriagulo}
\end{figure}

In general, the scalar particle production (decay) thorough VB fusion (emission) process 
\begin{equation*}
	V^\mu+V^\nu \longrightarrow\phi \mbox{\ \ (or \ \ }\phi \longrightarrow V^\mu +V^\nu),
\end{equation*}
is described by the invariant matrix element
\begin{equation}                                   
            \mathcal{M}\equiv\mathcal{M}^{\mu\nu}\epsilon_\mu\left(p_1,\lambda_1\right)\epsilon_\nu\left(p_2,\lambda_2\right),
    \label{eq.vacio-amplitud}
\end{equation}
where $p_i$ and $\lambda_i$, with $i=1,2$, are the momenta and polarization states of the incoming VB and $w$ is the scalar boson momentum.  $\epsilon_\mu\left(p_i,\lambda_i\right)$ are the polarization vectors that describe the asymptotic VB states. $\mathcal{M}^{\mu\nu}$ encodes the overlap between the initial and final states. In the model given by Eq.~(\ref{eq.lagrangian}), the leading order contributions to $\mathcal{M}^{\mu\nu}$ come from the type of Feynman diagram shown in Fig.\ref{figuratriagulo}.

The amplitudes $\mathcal{M}$, for the emission and fusion processes, share the same structure since are connected to each other by means of cross symmetry. From the invariant matrix elements we can build two physical observables that allows to quantify each process: the cross section and the decay width.

The differential cross section for the scalar boson production through VB fusion, $V^\mu+V^\nu\longrightarrow\phi$, is defined as~\cite{ParticleReview}
\begin{equation}
d\sigma=\dfrac{\sum_{\text{spin}}|\mathcal{M}|^2}{4\sqrt{\left(p_1\cdot p_2\right)^2}}\dfrac{d^3w}{(2\pi)^3 2E_w}(2\pi)^4\delta^{(4)}\left(p_1+p_2-w\right),
	\label{eq.sinB-diferentialcrosssection}
\end{equation}
where $E_w$ refers to the energy of the scalar boson.

On the other hand, the differential scalar boson decay width into two VB, $\phi\longrightarrow V^\mu+V^\nu$, is given by~\cite{ParticleReview}
\begin{equation}
	\begin{split}
	   d\Gamma&=\dfrac{\sum_{\text{spin}}|\mathcal{M}|^2}{2E_w}\dfrac{d^3p_1}{(2\pi)^3 2E_{1}}\dfrac{d^3p_2}{(2\pi)^3 2E_{2}}(2\pi)^4\delta^{(4)}\left(w-p_1-p_2\right),
	\end{split}
	\label{eq.sinB-diferentialdecayrate}
\end{equation}
where $E_i$ refers to the VB energies.

The factors along with $\mathcal{M}$, that refers to the final states kinematics, depend on the particle's nature and are independent of the interaction. Hence, all the information about the interaction is fully encoded in the amplitude.

In the particular case of a decaying process, $\phi\rightarrow V^\mu + V^\nu$ (calculated in Refs.~\cite{Spira,Marciano2012a} in the Higgs physics context), the amplitude has the form\footnote{We adapted the coupling constants to our model leaving the rest of the notation as appears in Ref.~\cite{Marciano2012a}.}
\begin{equation}
     \mathcal{M}(H\rightarrow\gamma\gamma)
     =\frac{ g^2 h}{(4\pi)^4m_W}A_f(\tau_f)\left(p_1\cdot p_2 \  g^{\mu\nu}-p_2^\mu p_1^\nu\right)\epsilon_\mu(p_1)\epsilon_\nu(p_2),
\label{marciano}
\end{equation}
where the factor $A_f(\tau)$, given by
\begin{equation}
     A_f(\tau) = 2[\tau+(\tau-1)f(\tau)]/\tau^2,
     \label{marciano2}
\end{equation}
accounts for the quantum fluctuations coming from fermion loops, with  $\tau=4m_f^2/m_H^2$ and 
\begin{equation}
   f(\tau)=\left\{\begin{array}{ll}
                            \arcsin^2(\tau^{-1/2}) & \mbox{for} \ \tau \   \geq 1 \\
                            -\frac{1}{4}\left(\ln\frac{1+\sqrt{1-\tau}}{1-\sqrt{1-\tau}}\right)^2 & \mbox{for} \ \tau <1
                       \end{array}
              \right. .
              \label{marciano3}
\end{equation}

\begin{figure}[H]
	\centering
	\includegraphics[width=4cm]{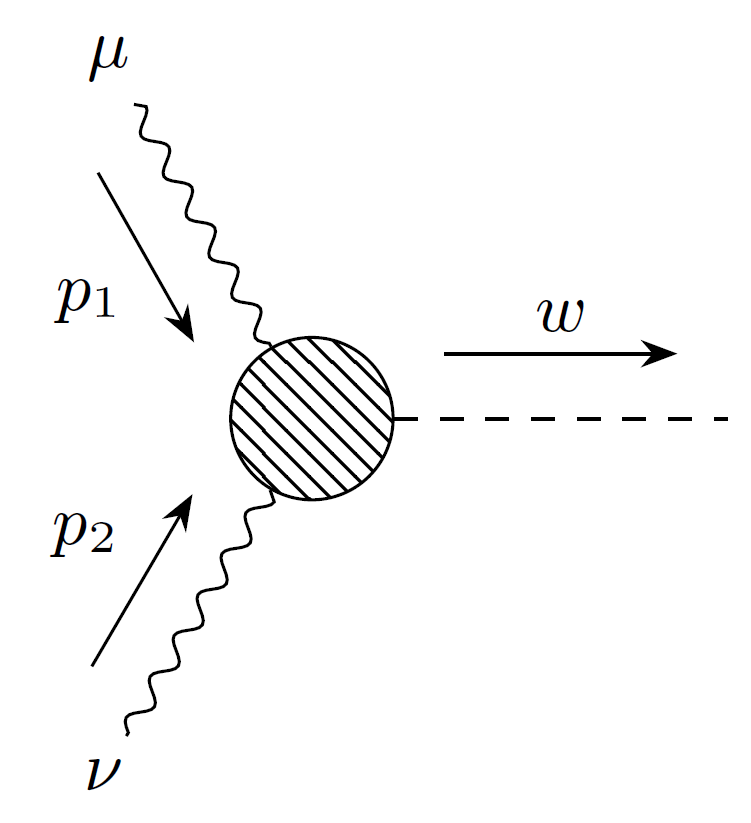}
\caption{Effective vertex that accounts the interaction between a scalar and two VB fields.}
	\label{fig.verticeefectivo}
\end{figure}

As the amplitude in Eq.~(\ref{marciano}) describes the interaction's strength among these three particles, the overlap in Eq.~(\ref{eq.vacio-amplitud}) could be treated as an effective vertex and pictorially associate with the Feynman diagram displayed in Fig.\ref{fig.verticeefectivo}. Since we are interested in a physical scenario where an external magnetic field is present, we expect that the interaction strength be modified. The computation of the effective vertex, and hence the amplitude, in presence of an external magnetic field is the main purpose of this work.

Based on general grounds, as symmetries and particle properties, in the next section we present the ideas behind the tensor structure in Eq.~(\ref{marciano}). This will allow us to build the tensor vertex structure in the presence of an external magnetic field in a simple way.

\section{Tensor vertex structure}
\label{sec.tensor}

\noindent In general, the effective vertex shown in Fig.~\ref{fig.verticeefectivo}, must satisfy 
\begin{equation}
	\label{eq.sinB-Slavnov1}
	p_1^\mu\mathcal{M}_{\mu\nu}(p_1,p_2)=0\mbox{\  \ and \ \ } 
	p_2^\nu\mathcal{M}_{\mu\nu}(p_1,p_2)=0,
\end{equation}
where $p_1^\mu$ and $p_2^\nu$ are the momenta for the incoming VB. In QED, the above relations are the well-known Ward identities~\cite{Ward,Taka}, which emerge from the current conservation and assure the effective vertex transversality to the VB momenta.

In addition, due to the indistinguishability between the incoming VB, the effective vertex must be invariant to the  vectorial boson exchange, it is
\begin{align}
	\mathcal{M}^{\mu\nu}(p_1,p_2)=\mathcal{M}^{\nu\mu}(p_2,p_1).
	\label{eq.sinB-PropBosones}
\end{align}
Besides these properties, the vertex must be also invariant under the discrete transformations of charge conjugation $(C)$ and parity $(P)$~\cite{QEDlandau}. Note that the effective vertex must fulfill the above properties regardless if the particles are in vacuum or in the presence of an external magnetic field. In what follows, we analyze its general tensor form in both scenarios.

\subsection{Vacuum vertex structure}
\label{subsec.tensorvac}

\noindent Let us start by considering that the tensors, in vacuum, at our disposal are
\begin{equation}
	p_1^\mu\text{, }p_2^\mu\text{, }g^{\mu\nu}\text{ and }\epsilon^{\mu\nu\alpha\beta}, 
\label{Vactens}	
\end{equation}
with $p_{i}^\mu$ the vector boson momenta, $g^{\mu\nu}=diag(+,-,-,-)$ the metric tensor and $\epsilon^{\mu\nu\alpha\beta}$ the Levi-Civita tensor. With these tensors, the most general form for the effective vertex, reads 
\begin{equation}
	\mathcal{M}_\text{vac.}^{\mu\nu}(p_1,p_2)=A_1g^{\mu\nu}+A_2p_1^\mu p_2^\nu+A_3p_1^\nu p_2^\mu+A_4p_1^\mu p_1^\nu+A_5p_2^\mu p_2^\nu+A_6\epsilon^{\mu\nu\alpha\beta}p_{1\alpha}p_{2\beta},
	\label{eq.sinB-EstructuraVertice}
\end{equation}
where coefficients depend on the different Lorentz scalars obtained from Eq.~(\ref{Vactens}). Note that, for a parity-conserved theory the coefficient $A_6$ vanishes identically. 

In the case where the incoming VB are on-shell ($p_1^2=p_2^2=0$) and the properties in Eqs.~(\ref{eq.sinB-Slavnov1}, \ref{eq.sinB-PropBosones}) are demanded to Eq.~(\ref{eq.sinB-EstructuraVertice}), the tensor structure reduces to
\begin{equation}
	\mathcal{M}_\text{vac.}^{\mu\nu}(p_1,p_2)=A_1\left[g^{\mu\nu}-\dfrac{p_1^\nu p_2^\mu}{p_1\cdot p_2}\right]+A_2p_1^\mu p_2^\nu, 
	\label{eq.sinB-EstructuraVerticeFinal}
\end{equation}
where the first term has the same tensor structure shown in Eq.~(\ref{marciano}).  Note that the second term in Eq.~(\ref{eq.sinB-EstructuraVerticeFinal}) does not contribute to the amplitude due to $p_i^\mu\epsilon_\mu(p_i)=0$. An important feature in Eq.~(\ref{eq.sinB-EstructuraVerticeFinal}) is that the two tensor structures are orthogonal one to each other.

\subsection{Magnetic field vertex structure}
\label{subsec.tensormag}

\noindent In the presence of a constant external magnetic field, described by the field strength tensor $F^{\mu\nu}$, we need to take into account that the number of independent four-vectors becomes increased to eight~\cite{Batalin,RITUS,Hugo}, namely
\begin{equation}
	p_i^\mu,\ F^{\mu\nu}{p_i}_{\nu}, \ {F^\mu}_\alpha F^{\alpha\nu}{p_i}_{\nu} \ \text{  and  } \ F^{*\mu\nu}{p_i}_{\nu},
\label{newvectors}
\end{equation}
where $i=1,2$, with 
\begin{equation}
	F^*_{\mu\nu}\equiv\frac{1}{2}\epsilon_{\mu\nu\gamma\delta}F^{\gamma\delta},
\end{equation}
the dual electromagnetic field strength tensor, which, in the present case, satisfies the cross field condition, $F^{\mu\nu}F^*_{\mu\nu}$=0.

Following Ref.~\cite{Batalin}, to build orthogonal tensor structures with the above vectors, let us consider the next complete set of orthogonal vectors\footnote{By complete, we mean
\begin{equation*}
	g^{\mu\nu}=\dfrac{l_{i}^{\mu}l_{i}^{\nu}}{l_{i}^{2}}+\dfrac{L_{i}^{\mu}L_{i}^{\nu}}{L_{i}^{2}}+\dfrac{L_{i}^{*\mu}L_{i}^{*\nu}}{L_{i}^{*2}}+\dfrac{G_{i}^{\mu}G_{i}^{\nu}}{G_{i}^{2}},
\end{equation*}
for $i=1,2$.}
\bea
	l^{\mu}_i\equiv p_i^{\mu}, \  \
	L^{\mu}_i\equiv \hat{F}^{\mu\nu}{p_i}_{\nu},\  \
	L^{*\mu}_i\equiv \hat{F}^{*\mu\nu}{p_i}_{\nu}\ \ \mbox{ and } \
	G^{^\mu}_i\equiv \frac{l^2}{L^2}\hat{F}^{\mu\alpha}\hat{F}_{\alpha\beta}p_i^{\beta}+l_i^{\mu},
\label{setspolvects}
\eea 
where $\hat{F}^{\mu\nu}\equiv F^{\mu\nu}/|B|$. In the above equation, the last three vectors can be used to describe the VB polarization states (for more details, see Appendix \ref{ap.polarizacion}).

Next, taking into account that each Lorentz index describe a VB with given momentum, it is not difficult to see that the most general form of the effective vertex, that fulfills Eq.~(\ref{eq.sinB-Slavnov1})~\cite{RitusPapan,RitusPapan2}, is
\begin{equation}
	\begin{split}
		\mathcal{M}_{{qB}}^{\mu\nu}(p_1,p_2)=&
   a_1^{++}\hat{L}_{1}^{\mu}\hat{L}_{2}^{\nu}
  +a_2^{++}\hat{L}_{1}^{*\mu}\hat{L}_{2}^{*\nu}
  +a_3^{++}\hat{G}_{1}^{\mu}\hat{G}_{2}^{\nu}
  +a_4^{+-}\hat{L}_{1}^{\mu}\hat{L}_{2}^{*\nu}\\
		&
  +a_5^{+-}\hat{L}_{1}^{*\mu}\hat{L}_{2}^{\nu}
  +a_6^{-+}\hat{L}_{1}^{\mu}\hat{G}_{2}^{\nu}
  +a_7^{-+}\hat{G}_{1}^{\mu}\hat{L}_{2}^{\nu}
  +a_8^{--}\hat{L}_{1}^{*\mu}\hat{G}_{2}^{\nu}
  +a_9^{--}\hat{G}_{1}^{\mu}\hat{L}_{2}^{*\nu},
	\end{split}
	\label{eq.MF.Esttensor1}
\end{equation}
where the superscripts ``$\pm$" in the  $a_k$ coefficients indicate its behavior under charge (first) and parity (second) transformations, respectively, and the ``hat" over vectors means they are normalized to the unit.  This structure is strongly linked to the Furry's theorem~\cite{Batalin} and resorts on the electromagnetic tensor field behavior under parity and charge transformations.

Finally, by requiring boson exchange symmetry in Eq.~(\ref{eq.sinB-PropBosones}) and considering on-shell VB, the effective vertex in presence of an external magnetic field, becomes\footnote{For on-shell VB, $G^{\mu}$  becomes proportional to $l^\mu$.}
\begin{equation}
	\mathcal{M}_{{qB}}^{\mu\nu}(p_1,p_2)=
       a_1^{++}\hat{L}_{1}^{\mu}\hat{L}_{2}^{\nu}
      +a_2^{++}\hat{L}_{1}^{*\mu}\hat{L}_{2}^{*\nu}
      +a_4^{+-}\dfrac{1}{\sqrt{2}}\left(\hat{L}_{1}^{\mu}\hat{L}_{2}^{*\nu}
      +\hat{L}_{1}^{*\mu}\hat{L}_{2}^{\nu}\right).
	\label{eq.MF.EsttensorOnShell}
\end{equation}

With the above orthogonal tensor decomposition for the effective vertex, we can easily compute the square amplitude, in Eqs.~(\ref{eq.sinB-diferentialcrosssection}, \ref{eq.sinB-diferentialdecayrate}), as (see Appendix \ref{ap.polarizacion})
\begin{equation}
\sum_{\text{spin}}|\mathcal{M}_{{qB}}|^2
=|a_1^{++}|^2+|a_2^{++}|^2+|a_4^{+-}|^2,
	\label{eq.sinB-PromEspin1}
\end{equation}
where each coefficient can be obtained by projecting the whole vertex with its corresponding tensor structure, it is
\begin{equation}
	\label{eq.coefa1}
	a_1^{++}=\mathcal{M}_{{qB}}^{\mu\nu}\hat{L}_{1\mu}\hat{L}_{2\nu},\ \
        a_2^{++}=\mathcal{M}_{{qB}}^{\mu\nu}\hat{L}_{1\mu}^{*}\hat{L}_{2\nu}^{*}\ \ \mbox{and} \ \
	a_4^{+-}=\mathcal{M}_{{qB}}^{\mu\nu}\dfrac{1}{\sqrt{2}}\left(\hat{L}_{1\mu}\hat{L}_{2\nu}^{*}+\hat{L}_{1\mu}^{*}\hat{L}_{2\nu}\right).
\end{equation}
Here, we can see that once the coefficient are obtained, by projecting the effective vertex onto the orthogonal basis, the physical observables can be computed straightforward. For example, replacing Eq.~(\ref{eq.sinB-PromEspin1}) in Eq.~(\ref{eq.sinB-diferentialcrosssection}) and performing the integration and the average over spins, the unpolarized cross section in the presence of a magnetic field can be obtained
\begin{equation}
	\sigma_{qB}\left(\phi\longrightarrow VV\right)=\dfrac{1}{8m_\phi^2}\left(|a_1^{++}|^2+|a_2^{++}|^2+|a_4^{+-}|^2\right)2\pi\delta\left(\mathcal{S}-m_\phi^2\right),
	\label{eq.sinB-crosssectionFinal}
\end{equation}
where $\mathcal{S}=(p_1+p_2)^2$ is the usual variable for the square energy, $m_\phi$ is the mass of the scalar boson and the delta function assures that it is produced on-shell.

So far, based on general grounds, we have written the effective vertex in terms of orthogonal tensor structures. However, we do not known the explicit form of the amplitude which encodes the microphysics that allows the interaction between these three particles. In general, up to a certain order of approximation, the computation of $\mathcal{M}_{{qB}}^{\mu\nu}$ (which requires a finite sum of Feynman diagrams) is not expressed in a closed form of orthogonal tensor structures as in Eq.~(\ref{eq.MF.EsttensorOnShell}). In the next section, we shall focus on the calculation of  the effective vertex up to one-loop fermion, left hand side ({\it lhs}) in Eq.~(\ref{eq.MF.EsttensorOnShell}).

\section{One-loop effective vertex}
\label{sec.calculations}

\noindent In order to incorporate the magnetic field effects on the one-loop effective vertex, shown in Fig.\ref{figuratriagulo}, we consider the fermion charged particles propagation in the presence of an external magnetic field, whose general structure reads~\cite{ERDAS}
\begin{equation}
	S^{^{qB}}(x,y)=\Omega(x,y)\int \dfrac{d^4p}{(2\pi)^4}\tilde{S}^{^{qB}}(p)e^{-ip\cdot(x-y)},
	\label{eq.conB-PropConf}
\end{equation}
where
\begin{equation}
	\Omega(x',x'')=\exp\left(-iq\int_{x''}^{x'} A_\mu(x)dx^\mu\right),
	\label{eq.conB-FaseSchwinger}
\end{equation}
is the well known Schwinger's phase, with $q$ the electric charge of the fermion and $A_\mu(x)$ the four-vector potential associated with the magnetic field. In the particular case of an homogeneous magnetic field along the $z$-direction ($F^{21}=-F^{12}=B$), the translationally invariant part of the fermionic propagator has the form~\cite{ERDAS}
\begin{equation}
	\begin{split}
		\tilde{S}^{^{qB}}(p)=\int_{0}^{\infty}&\dfrac{ds}{\cos(qBs)}\exp\bigg[-is\left(m^2-p_{\parallel}^2-p_{\perp}^2\dfrac{\tan(qBs)}{qBs}\right)\bigg]\\
		\times&\left[\left(m+\slsh{p}_{\parallel}\right)e^{iqBs\Sigma_3}+\dfrac{\slsh{p}_{\perp}}{\cos(qBs)}\right],
	\end{split}
	\label{eq.conB-PropMomento}
\end{equation}
where the parameter $s$ is the Schwinger's proper time and $qB=\text{sign}(qB)|qB|$. The momentum components $p_{\parallel}$ and $p_{\perp}$, parallel and perpendicular to the magnetic field, are given by
\begin{equation}
	p_{\parallel}^\mu\equiv(p^0,0,0,p^3) \ \mbox{ and } \ p_{\perp}^\mu\equiv(0,p^1,p^2,0), 
\end{equation}
satisfying $p^2=p_\parallel^2+p_\perp^2$. Finally, $m$ is the fermion mass and $\Sigma_3\equiv i\gamma^1\gamma^2$ is related with the spin along $z$-direction.

\begin{figure}[H]
	\centering
	\subfigure[Diagram I.]{\includegraphics[width=6.3cm]{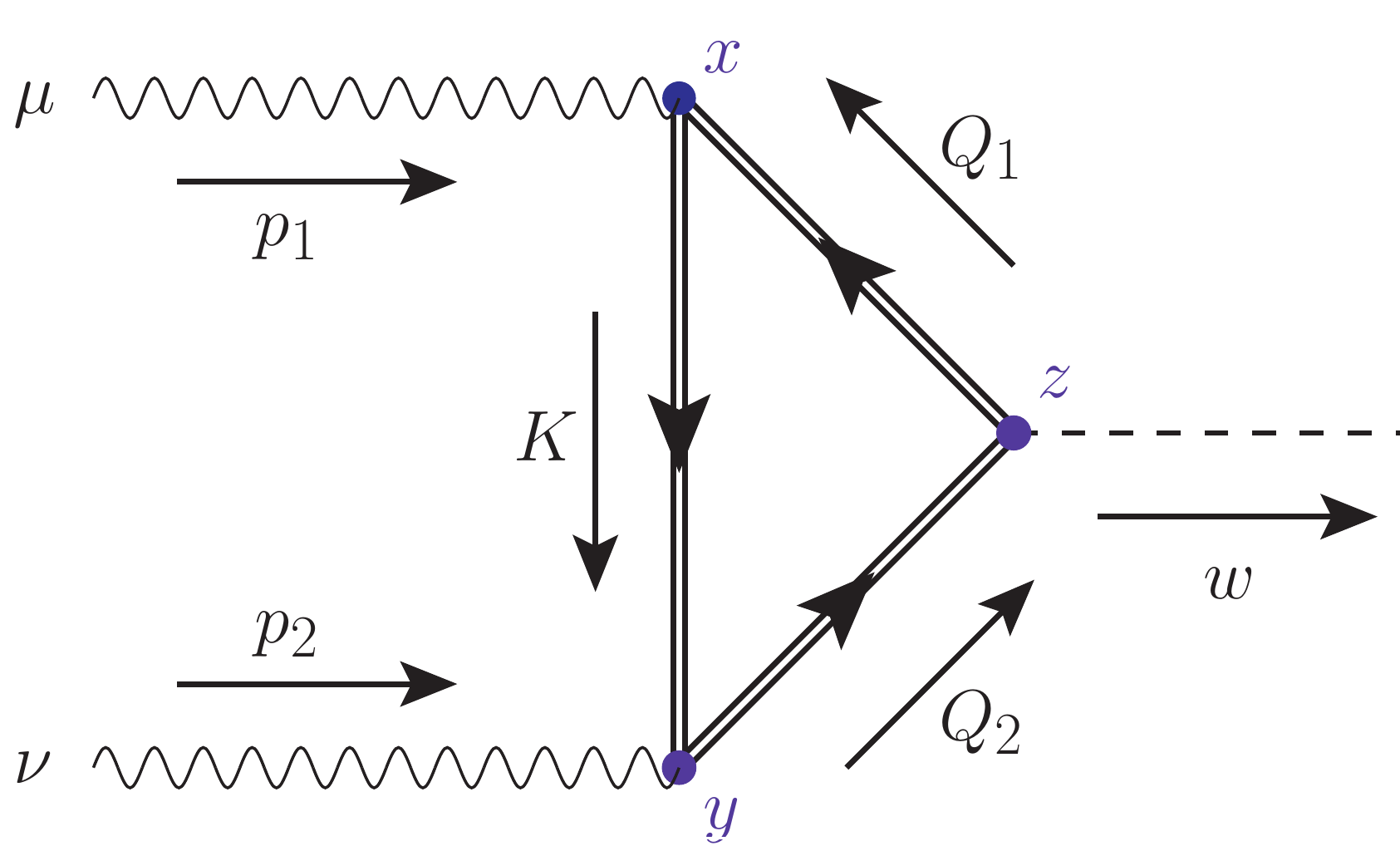} }%
	\qquad
	\subfigure[Diagram II.]{\includegraphics[width=6.3cm]{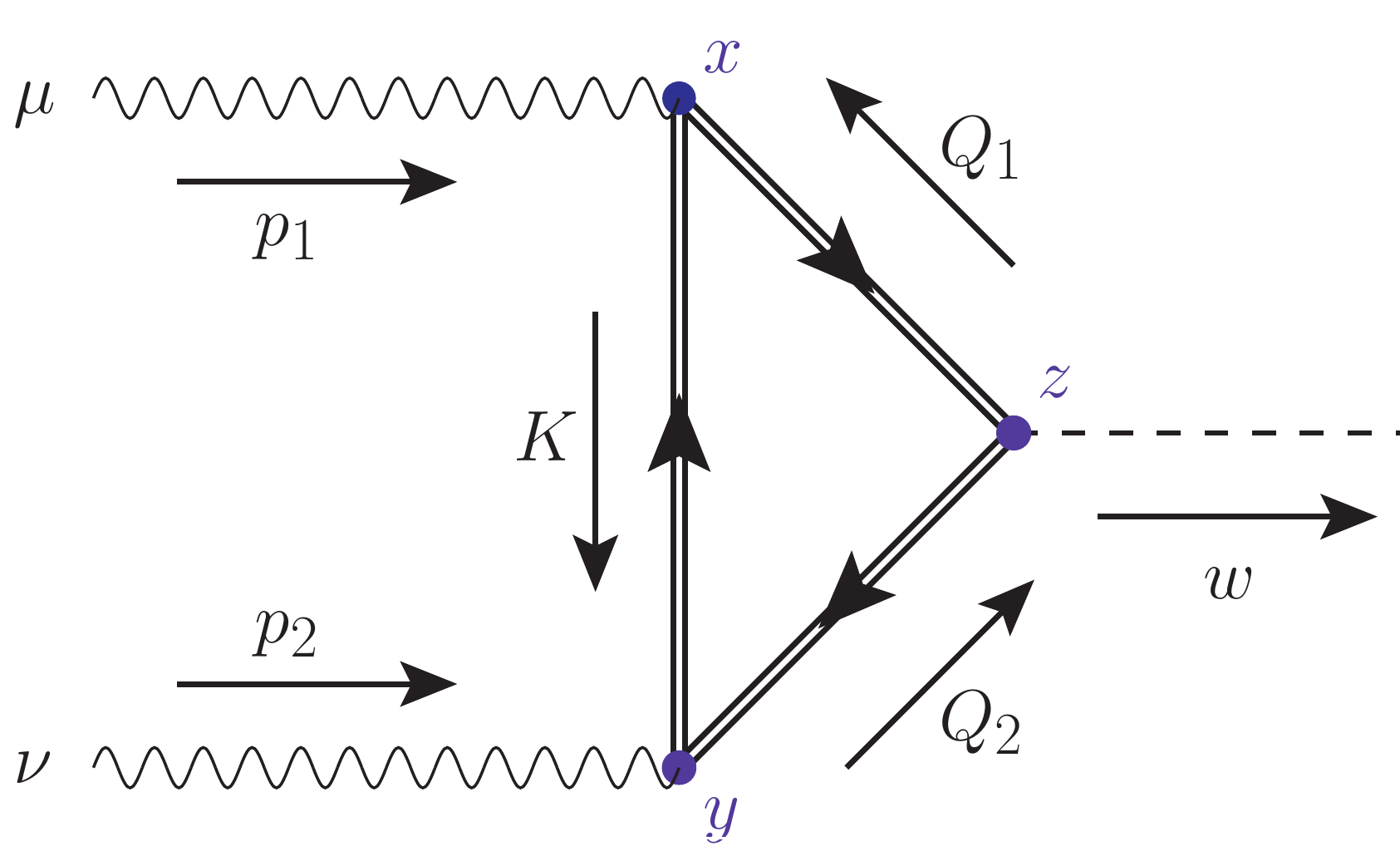} }%
	\caption{$VV\Phi$ one-loop Feynman diagrams in presence of an external magnetic field. The external magnetic field is indicated by the double line in the fermion propagators, the internal arrows indicate the charge flux and the external arrows the momentum flux. $x$, $y$ and $z$ indicate the interaction points in the configuration space.}
	\label{fig.diagramasChidosCampo}
\end{figure}

In Fig. \ref{fig.diagramasChidosCampo}, we show the two (charge conjugate) Feynman diagrams contributing to the process at one-loop in which the magnetic field effect is represented by a double line in the fermionic propagators. 

Applying the Feynman rules to the diagrams shown in Fig. \ref{fig.diagramasChidosCampo}, we get the following analytic expressions
\begin{equation}
		i{\mathcal{M}_{{qB}}^{\mu\nu}}_{(I)}(x,y,z)=-\mathcal{TR}\left[\left(-ig\gamma^\mu\right)S^{^{qB}}(x,z)\left(-ih\right)S^{^{qB}}(z,y)\left(-ig\gamma^\nu\right)S^{^{qB}}(y,x)\right],
	\label{eq.MF-VEdiagA}
\end{equation}
and
\begin{equation}
		i{\mathcal{M}_{{qB}}^{\mu\nu}}_{(II)}(x,y,z)=-\mathcal{TR}\left[\left(-ig\gamma^\mu\right)S^{^{qB}}(x,y)\gamma^\nu\left(-ig\gamma^\nu\right)S^{^{qB}}(y,z)\left(-ih\right)S^{^{qB}}(z,x)\right],
	\label{eq.MF-VEdiagB}
\end{equation}
where $g$ and $h$ are the corresponding interaction coupling constant between charged fermions with vector and scalar bosons, respectively. $\mathcal{TR}$ indicates sum over all internal degrees of freedom. 

Due to Eqs.~(\ref{eq.MF-VEdiagA}, \ref{eq.MF-VEdiagB}), are related by charge conjugation transformation, $\hat{\mathcal{C}}$, namely
\begin{equation}
	\hat{\mathcal{C}}{\mathcal{M}_{{qB}}^{\mu\nu}}_{(I)}\hat{\mathcal{C}}^{-1}
 ={\mathcal{M}_{{qB}}^{\mu\nu}}_{(II)},
    \label{eq.chargueconj}
\end{equation}
from now on we focus our analysis to diagram I.

The effective vertex in the momentum space is obtained as follows
\begin{equation}
	{\mathcal{M}_{{qB}}^{\mu\nu}}(p_1,p_2,w)=\int d^Dx\hspace{1mm}d^Dy\hspace{1mm}d^Dz\hspace{1mm} {\mathcal{M}_{{qB}}^{\mu\nu}}(x,y,z) e^{-ip_1\cdot x}e^{-ip_2\cdot y}e^{+iw\cdot z}, 
	\label{eq.MF-VEdiagFourier}
\end{equation}
where we have extended the space-time to $D$-dimensions to deal with the logarithmic divergences that comes from the vacuum part in Eqs.~(\ref{eq.MF-VEdiagA}, \ref{eq.MF-VEdiagB}). Notice that the plane waves are used due to the neutral nature of the external particles.

Once we replace Eq.~(\ref{eq.conB-PropConf}) in Eq.~(\ref{eq.MF-VEdiagA}), the diagram I in momentum space, reads
\begin{equation}
	\begin{split}
		i{\mathcal{M}_{{qB}}^{\mu\nu}}_{(I)}(p_1,p_2,w)=&-ih g^2\int\dfrac{d^DK\hspace{1mm}d^DQ_1\hspace{1mm}d^DQ_2}{(2\pi)^{D}(2\pi)^{D}(2\pi)^{D}}\\
		&\times\int d^Dx\hspace{1mm}d^Dy\hspace{1mm}d^Dz\hspace{1mm}e^{-i(p_1+Q_1-K)\cdot x}e^{-i(p_2+K-Q_2)\cdot y}e^{-i(-w+Q_2-Q_1)\cdot z}\\
		&\hspace{30mm}\times e^{i\frac{qB}{2}\left(x\hat{F}z+z\hat{F}y+y\hat{F}x\right)}\textit{Tr}\left[\gamma^\mu\tilde{S}^{^{qB}}(Q_1)\tilde{S}^{^{qB}}(Q_2)\gamma^\nu\tilde{S}^{^{qB}}(K)\right],
	\end{split}
	\label{eq.MF-VEdiagA2.}
\end{equation}
where $Tr$ denotes sum over spin and we adopted the notation $a\hat{F}b\equiv a_\mu \hat{F}^{\mu\nu} b_\nu$. Note that in the above equation, the exponential factor, next to $Tr$,  emerges from the Schwinger's phases in each fermion propagator (for more details see Appendix~\ref{ap.fase}). 

In what follows, we show a novel a methodology that allow us to obtain a close analytical expression for the effective vertex, Eq.~(\ref{eq.MF-VEdiagA2.}), in an homogeneous magnetic field with arbitrary strength.

\subsection{Coordinate space integration}

\noindent Let us start by performing the integrals over the coordinate space. Due the presence of the magnetic field two kinds of integrals emerge, associated to the parallel and perpendicular components. The integration over the parallel coordinate space in Eq.~(\ref{eq.MF-VEdiagA2.}) can be performed straightforward, giving
\begin{equation}
	\begin{split}
		i{\mathcal{M}_{{qB}}^{\mu\nu}}_{(I)}(p_1,p_2,w)=&-ih g^2\int\dfrac{d^DK\hspace{1mm}d^{D}Q_{1}\hspace{1mm}d^{D}Q_{2}}{(2\pi)^{D}(2\pi)^{D}(2\pi)^{D}}(2\pi)^{D_\parallel}\delta^{(D_\parallel)}_\parallel\left(p_1+Q_1-K\right)\\
		&\times(2\pi)^{D_\parallel}\delta^{(D_\parallel)}_\parallel\left(p_2+K-Q_2\right)(2\pi)^{D_\parallel}\delta^{(D_\parallel)}_\parallel\left(-w+Q_2-Q_1\right)\\
		&\times \textit{Tr}\left[\gamma^\mu\tilde{S}^{^{qB}}\left(Q_{1\perp},Q_{1\parallel}\right)\tilde{S}^{^{qB}}\left(Q_{2\perp},Q_{2\parallel}\right)\gamma^\nu\tilde{S}^{^{qB}}\left(K_{\perp},K_\parallel\right)\right]\\
		&\times\int d^{D_\perp}x_{\perp}\hspace{1mm}d^{D_\perp}y_{\perp}\hspace{1mm}d^{D_\perp}z_{\perp}\hspace{1mm}e^{-i(p_1+Q_1-K)_{\perp}\cdot x_{\perp}}e^{-i(p_2+K-Q_2)_{\perp}\cdot y_{\perp}}\\
		&\hspace{45mm}\times e^{-i(-w+Q_2-Q_1)_{\perp}\cdot z_{\perp}} e^{i\frac{qB}{2}\left(x\hat{F}z+z\hat{F}y+y\hat{F}x\right)}.
	\end{split}
	\label{eq.MF-VEdiagA2}
\end{equation}
From the above equation, it is easily to observe the parallel momenta conservation in each vertex. Note that the Schwinger phase exponential factor was unaffected because the contractions $a\hat{F}b$ only admits perpendicular components. 

On the other hand, the integration over the perpendicular components is a little bit more involved. To carry out this integration in Eq.~(\ref{eq.MF-VEdiagA2}), let us isolate the following structure 
\begin{equation}
		I_\perp\equiv\int d^{D_\perp}x_{\perp}\hspace{1mm}d^{D_\perp}y_{\perp}\hspace{1mm}d^{D_\perp}z_{\perp}\hspace{1mm}e^{-i(p_1+Q_1-K)_{\perp}\cdot x_{\perp}}e^{-i(p_2+K-Q_2)_{\perp}\cdot y_{\perp}}e^{-i(-w+Q_2-Q_1)_{\perp}\cdot z_{\perp}} e^{i\frac{qB}{2}\left(x\hat{F}z+z\hat{F}y+y\hat{F}x\right)}.
\end{equation}

Due to the coordinate mixing that appears in the Schwinger phase exponential factor, it is not possible to identify Dirac's delta functions as in the previous case. This factor breaks down the perpendicular momenta conservation in each vertex by mixing their interaction points. However, the integration can be done (component by component), and once is completed the perpendicular momenta conservation, for the overall process, becomes explicit.

For example, in the first step, the integration over $x_\perp$ give rises to a Dirac delta function with an argument that depends on $y_\perp$ and $z_\perp$. In the second step, the integration over $y_\perp$ is straightforward because the previous a Dirac's delta function. Finally, the integral over $z_\perp$ is again identified as a Dirac delta function, whose argument shows the perpendicular momenta conservation, getting
\begin{equation}
	I_\perp=\left(\frac{4\pi}{|qB|}\right)^{D_\perp}(2\pi)^{D_\perp}\delta^{(D_\perp)}_\perp\left(p_1+p_2-w\right) e^{i\frac{2}{qB}(p_1+Q_1-K)\hat{F}(p_2+K-Q_2)},
\label{eq.fase1}
\end{equation}
where the exponential factor comes from the Schwinger's phases. Different orders of integration over $x_\perp$, $y_\perp$ and $z_\perp$ gives different results which seems to depend on the order of integration. 
However, when integration over the loop momenta is carried out, the final result is the same regardless of the integration order in coordinate space.

Using the result given by Eq.~(\ref{eq.fase1}) in Eq.~(\ref{eq.MF-VEdiagA2}), we get
\begin{equation}
	\begin{split}
		i{\mathcal{M}_{{qB}}^{\mu\nu}}_{(I)}(p_1,p_2,w)
        =&-ih g^2\left(\frac{4\pi}{|qB|}\right)^{D_\perp}(2\pi)^{D}\delta^{(D)}\left(p_1+p_2-w\right)\\
		&\times\int\dfrac{d^D K\hspace{1mm}d^{D_\perp}Q_{1\perp}\hspace{1mm}d^{D_\perp}Q_{2\perp}}{(2\pi)^{D}(2\pi)^{D_\perp}(2\pi)^{D_\perp}}e^{i\frac{2}{qB}(p_1+Q_1-K)\hat{F}(p_2+K-Q_2)}\\
		&\times \textit{Tr}\left[\gamma^\mu\tilde{S}^{^{qB}}\left(Q_{1\perp},(K-p_1)_\parallel\right)\tilde{S}^{^{qB}}\left(Q_{2\perp},(K+p_2)_\parallel\right)\gamma^\nu\tilde{S}^{^{qB}}\left(K_{\perp},K_\parallel\right)\right]\\
  \equiv&-ih g^2\left(\frac{4\pi}{|qB|}\right)^{D_\perp}(2\pi)^{D}\delta^{(D)}\left(p_1+p_2-w\right)G_{\left(I\right)}.
	\end{split}
	\label{eq.MF-VEdiagA3}
\end{equation}
where the integration over $Q_{1\parallel}$ and $Q_{2\parallel}$ has been additionally done and contributes to the full momentum conservation factor that, for simplicity, will be omitted from now on. $G_{(I)}$ was introduced in advance to work with the momenta integrals in the next part.

So far, we have finished with the coordinate space integration and carried out the trivial momentum integration in advance. There are four momentum integrals left to be performed, three perpendicular and one parallel, which require a more exhaustive procedure. In what follows, we present a methodology to deal with this kind of integration.

\subsection{Momentum integration}

\noindent In a standard calculation procedure of Eq.~(\ref{eq.MF-VEdiagA3}), the next step is to carry out the sum over spins in the loop.  This step gives rise to a huge amount of terms which become really difficult to deal with. Instead of the standard procedure, we shall first calculate the integrals over all momenta, involved in the loop, leaving at the end the sum over spins. With this goal in mind, let us write the expression we are going to work on
\begin{equation}
	\begin{split}
		G_{(I)}=\int\dfrac{ds_1\hspace{1mm}ds_2\hspace{1mm}ds_3}{\mathrm{c_1}^2\mathrm{c_2}^2\mathrm{c_3}^2}\int&\dfrac{d^DK\hspace{1mm}d^{D_\perp}Q_{1\perp}\hspace{1mm}d^{D_\perp}Q_{2\perp}}{(2\pi)^{D}(2\pi)^{D_\perp}(2\pi)^{D_\perp}}e^{i\frac{2}{qB}(p_1+Q_1-K)\hat{F}(p_2+K-Q_2)}\\
		&\times e^{is_1\left(\left(K-p_1\right)_\parallel^2-m^2\right)+is_2\left(\left(K+p_2\right)_\parallel^2-m^2\right)+is_3\left(K_\parallel^2-m^2\right)}e^{i\frac{\mathrm{t_1}}{qB}Q_{1\perp}^2+i\frac{\mathrm{t_2}}{qB}Q_{2\perp}^2+i\frac{\mathrm{t_3}}{qB}K_{\perp}^2}\\
        &\hspace{45mm}\times\textit{Tr}\left[\gamma^\mu\left(\mathbb{M}_1+\slsh{Q}_{1\perp}\right)\left(\mathbb{M}_2+\slsh{Q}_{2\perp}\right)\gamma^\nu\left(\mathbb{M}_3+\slsh{K}_{\perp}\right)\right],
	\end{split}
	\label{eq.MF-VEdiagAintgauss2}
\end{equation}
where we have used the fermion propagators in Eq.~(\ref{eq.conB-PropMomento}) and introduced the notation
\begin{align}
	\label{eq.GaussianIntM1}
	\mathbb{M}_1&\equiv\left(m+(\slsh{K}-\slsh{p}_1)_\parallel\right)e^{(1)},\\
	\label{eq.GaussianIntM2}
	\mathbb{M}_2&\equiv\left(m+(\slsh{K}+\slsh{p}_2)_\parallel\right)e^{(2)},\\
	\mathbb{M}_3&\equiv\left(m+\slsh{K}_\parallel\right)e^{(3)},
	\label{eq.GaussianIntM3}
\end{align}
which emphasize the matricial nature of the quantities, with 
\begin{equation}
	e^{(j)}\equiv \mathrm{c_j}e^{iqBs_j\Sigma_3},
\end{equation}
and $\mathrm{c_j}\equiv\cos\left(qBs_j\right)$, $\mathrm{t_j}\equiv\tan\left(qBs_j\right)$ with $s_j$ the Schwinger's parameters of each fermion propagator. In Eq.~(\ref{eq.MF-VEdiagAintgauss2}), we separated the perpendicular and parallel components because the associated integrals are independent one to each other.

To perform the integration over one of the perpendicular components, for example the variable $Q_{1\perp}$, note that its integral
\begin{equation}
	G_{(I)}^{^{Q_{1\perp}}}\equiv\int\dfrac{d^{D_\perp}Q_{1\perp}}{(2\pi)^{D_\perp}}e^{i\left(\frac{\mathrm{t_1}}{qB}Q_{1\perp}^2+\frac{2}{qB}Q_1\hat{F}(p_2+K-Q_2)\right)}\textit{Tr}\left[\gamma^\mu\left(\mathbb{M}_1+\slsh{Q}_{1\perp}\right)\left(\mathbb{M}_2+\slsh{Q}_{2\perp}\right)\gamma^\nu\left(\mathbb{M}_3+\slsh{K}_{\perp}\right)\right],
	\label{eq.MF-VEdiagAintgauss3}
\end{equation}
has as a Gaussian form which, by making the change of variable
\begin{equation}
	l^\mu_{{Q_{1\perp}}}\equiv Q_{1\perp}^\mu-\frac{1}{\mathrm{t_1}}\hat{F}^{\mu\nu}\left(Q_2-K-p_2\right)_{\perp\nu},
	\label{eq.GaussianShiftQ1perp}
\end{equation}
can be rewritten as
\begin{equation}
	\begin{split}
		G_{(I)}^{^{Q_{1\perp}}}=&e^{-\frac{i}{qB\mathrm{t_1}}\left(Q_2-K-p_2\right)^2_\perp}\int\dfrac{d^{D_\perp}l_{{Q_{1\perp}}}}{(2\pi)^{D_\perp}}e^{i\frac{\mathrm{t_1}}{qB}{l^2_{Q_{1\perp}}}}\\
		&\times\textit{Tr}\left[\gamma^\mu\left(\mathbb{M}_1+\slsh{l}_{{Q_{1\perp}}}+\frac{1}{\mathrm{t_1}}\slsh{\hat{F}}\left(Q_2-K-p_2\right)_\perp\right)\left(\mathbb{M}_2+\slsh{Q}_{2\perp}\right)\gamma^\nu\left(\mathbb{M}_3+\slsh{K}_{\perp}\right)\right],
	\end{split}
	\label{eq.MF-VEdiagAintgauss4}
\end{equation}
where the price to write the exponential argument as a single quadratic term is paid by giving rise to extra terms in the spinorial trace that has the form $\slsh{\hat{F}}a=\gamma^\mu\hat{F}_{\mu\nu}a^\nu$. A relevant remark is that the ``shift'' comes from the Schwinger's phase exponential factor.

In the above expression, the spinorial trace is a polynomial function of order 1 on the variable $l_{Q_{1\perp}}$. By using the generalized result of Gaussian integration and symmetry arguments, it can be shown that the linear term vanishes, giving as a result
\begin{equation}
	\begin{split}
		G_{(I)}^{^{Q_{1\perp}}}=&e^{-\frac{i}{qB\mathrm{t_1}}\left(Q_2-K-p_2\right)^2_\perp}\dfrac{1}{(2\pi)^{D_\perp}}\left(\sqrt{\frac{-i\pi qB}{\mathrm{t_1}}}\right)^{D_\perp}\\
		&\hspace{13mm}\times\textit{Tr}\left[\gamma^\mu\left(\mathbb{M}_1+\frac{1}{\mathrm{t_1}}\slsh{\hat{F}}\left(Q_2-K-p_2\right)_\perp\right)\left(\mathbb{M}_2+\slsh{Q}_{2\perp}\right)\gamma^\nu\left(\mathbb{M}_3+\slsh{K}_{\perp}\right)\right].
	\end{split}
	\label{eq.MF-VEdiagAintgauss4b}
\end{equation}

Once we replace Eq.~(\ref{eq.MF-VEdiagAintgauss4b}) in Eq.~(\ref{eq.MF-VEdiagAintgauss2}), we arrive at
\begin{equation}
	\begin{split}
		G_{(I)}=\int&\dfrac{ds_1\hspace{1mm}ds_2\hspace{1mm}ds_3}{\mathbf{c_1}^2\mathbf{c_2}^2\mathbf{c_3}^2}\dfrac{1}{(2\pi)^{D_\perp}}\left(\sqrt{\frac{-i\pi qB}{\mathrm{t_1}}}\right)^{D_\perp}\\
		&\times\int\dfrac{d^DK}{(2\pi)^{D}}e^{is_1\left(\left(K-p_1\right)_\parallel^2-m^2\right)+is_2\left(\left(K+p_2\right)_\parallel^2-m^2\right)+is_3\left(K_\parallel^2-m^2\right)}\\
		&\hspace{14mm}\times e^{i\frac{\mathrm{t_3}}{qB}K_{\perp}^2}e^{i\frac{2}{qB}\left(p_1-K\right)\hat{F}\left(p_2+K\right)}e^{-\frac{i}{\mathrm{t_1}qB}\left(K+p_2\right)^2_\perp}\\
		&\hspace{14mm}\times\int\dfrac{d^{D_\perp}Q_{2\perp}}{(2\pi)^{D_\perp}}e^{\frac{i}{qB}\left(\mathrm{t_2}-\frac{1}{\mathrm{t_1}}\right)Q_{2\perp}^2}e^{i\frac{2}{qB}\left((p_1-K)\hat{F}Q_2+\frac{1}{\mathrm{t_1}}\left(K+p_2\right)_\perp\cdot Q_{2\perp}\right)}\\
		&\hspace{24mm}\times\textit{Tr}\left[\gamma^\mu\left(\mathbb{M}_1+\frac{1}{\mathrm{t_1}}\slsh{\hat{F}}\left(Q_2-K-p_2\right)_\perp\right)\left(\mathbb{M}_2+\slsh{Q}_{2\perp}\right)\gamma^\nu\left(\mathbb{M}_3+\slsh{K}_{\perp}\right)\right].
	\end{split}
	\label{eq.MF-VEdiagAintgauss5}
\end{equation}
In the above equation, since the functional structure on ${Q_{2\perp}}$ is quite similar as the one for ${Q_{1\perp}}$ in Eq.~(\ref{eq.MF-VEdiagAintgauss3}), the integration can be performed with roughly the same procedure: by making a ``shift'' that allows us to write the exponential argument as a single quadratic term\footnote{The ``shift'' is a little bit more involved that in the previous case since it acquires contributions from the Schwinger's exponential factor and from the integration over $Q_{1\perp}$.} and by noting that the spinorial trace is a polynomial function of order 2 on the variable ${Q_{2\perp}}$ which, by symmetry arguments, contribute to the integral with the quadratic and constant terms.

Following this line of thoughts, since the remaining integrals over $K_\perp$ and $K_\parallel$ share a similar functional structure as the ${Q_{1\perp}}$ and ${Q_{2\perp}}$ ones, they can be done by using the above-mentioned procedure. The difference is in the spinorial trace that will be a polynomial function of order 3 on such variables. In the general case, the trace will be a polynomial function of order $n$ on the integration variable, contributing to the integral only the even powers.

Once the full loop momenta integration is performed, according to the previous discussion, the final result reads
\begin{equation}
	\begin{split}
		i{\mathcal{M}_{{qB}}^{\mu\nu}}_{(I)}(p_1,p_2)=&ih g^2\left(-i\right)^{D_\perp+D/2+1}\dfrac{|qB|^{D_\perp/2}}{2^{D}\pi^{D/2}}\\
		&\times\int\dfrac{ds_1\hspace{1mm}ds_2\hspace{1mm}ds_3}{\mathrm{c_1}^2\mathrm{c_2}^2\mathrm{c_3}^2}\left(\frac{1}{s}\right)^{D_\parallel/2}\left(\frac{\text{sign}\left(qB\right)}{\mathrm{t_1}\mathrm{t_2}\mathrm{t_3}-\mathrm{t_1}-\mathrm{t_2}-\mathrm{t_3}}\right)^{D_\perp/2}e^{-ism^2}\\
		&\times e^{\frac{i}{s}\left(s_1s_3p_{1\parallel}^2+s_2s_3p_{2\parallel}^2+s_1s_2w_{\parallel}^2\right)}e^{-\frac{i}{qB}\frac{\mathrm{t_1}\mathrm{t_3}p_{1\perp}^2+\mathrm{t_2}\mathrm{t_3}p_{2\perp}^2+\mathrm{t_1}\mathrm{t_2}w_{\perp}^2+2\mathrm{t_1}\mathrm{t_2}\mathrm{t_3}p_2\hat{F}p_1}{\mathrm{t_1}\mathrm{t_2}\mathrm{t_3}-\mathrm{t_1}-\mathrm{t_2}-\mathrm{t_3}}}\\
		&\times\Bigg\{\textit{Tr}\left[\gamma^\mu\slsh{\mathbb{U}}_1\slsh{\mathbb{U}}_2\gamma^\nu\slsh{\mathbb{U}}_3\right]+\frac{im}{2s}\Bigg(-D_\parallel\textit{Tr}\left[\gamma^\mu e^{(1)}e^{(2)}\gamma^\nu e^{(3)}\right]\\
		&\hspace{60mm}+2\textit{Tr}\left[\gamma^\mu_\parallel e^{(1)}e^{(2)}\gamma^\nu e^{(3)}\right]\\
		&\hspace{66mm}+2\textit{Tr}\left[\gamma^\mu e^{(1)}e^{(2)}\gamma^\nu_\parallel e^{(3)}\right]\Bigg)\\
		&\hspace{9mm}+\frac{im \ qB }{2\left(\mathrm{t_1}\mathrm{t_2}\mathrm{t_3}-\mathrm{t_1}-\mathrm{t_2}-\mathrm{t_3}\right)}\Bigg(-D_\perp\textit{Tr}\left[\gamma^\mu\gamma^\nu e^{(3)}\right]\\
		&\hspace{56mm}+D_\perp\textit{Tr}\left[\gamma^\mu\left(e^{(1)}+e^{(2)}\right)\gamma^\nu\right]\\
		&\hspace{58mm}-2\textit{Tr}\left[\gamma^\mu e^{(1)}\gamma^\nu_\perp\right]-2\textit{Tr}\left[\gamma^\mu_\perp e^{(2)}\gamma^\nu\right]\\
		&\hspace{60mm}+\mathrm{t_1}\textit{Tr}\left[\gamma^\mu e^{(1)}\left(\gamma^\alpha_\perp\hat{F}_{\alpha\beta}\right)\gamma^\nu\gamma^\beta_\perp\right]\\
		&\hspace{62mm}+\mathrm{t_2}\textit{Tr}\left[\gamma^\alpha_\perp\gamma^\mu\left(\hat{F}_{\alpha\beta}\gamma^\beta_\perp\right)e^{(2)}\gamma^\nu\right]\\
		&\hspace{64mm}+\mathrm{t_3}\textit{Tr}\left[\gamma^\mu\left(\gamma^\alpha_\perp\hat{F}_{\alpha\beta}\gamma^\beta_\perp\right)\gamma^\nu e^{(3)}\right]\Bigg)\Bigg\},
	\end{split}
	\label{eq.MF-VEdiagA4}
\end{equation}	
where
\begin{align}
	\slsh{\mathbb{U}}_1&=\left(m-\frac{s_3\slsh{p}_{1\parallel}+s_2\slsh{w}_{\parallel}}{s}\right)e^{(1)}+\frac{\mathrm{t_3}\slsh{p}_{1\perp}+\mathrm{t_2}\slsh{w}_{\perp}-\mathrm{t_2}\mathrm{t_3}\slsh{\hat{F}}p_2 }{\mathrm{t_1}\mathrm{t_2}\mathrm{t_3}-\mathrm{t_1}-\mathrm{t_2}-\mathrm{t_3}},\\
	\slsh{\mathbb{U}}_2&=\left(m+\frac{s_1\slsh{w}_{\parallel}+s_3\slsh{p}_{2\parallel}}{s}\right)e^{(2)}+\frac{-\mathrm{t_3}\slsh{p}_{2\perp}-\mathrm{t_1}\slsh{w}_{\perp}-\mathrm{t_1}\mathrm{t_3}\slsh{\hat{F}}p_1 }{\mathrm{t_1}\mathrm{t_2}\mathrm{t_3}-\mathrm{t_1}-\mathrm{t_2}-\mathrm{t_3}},\\
	\slsh{\mathbb{U}}_3&=\left(m+\frac{s_1\slsh{p}_{1\parallel}-s_2\slsh{p}_{2\parallel}}{s}\right)e^{(3)}+\frac{-\mathrm{t_1}\slsh{p}_{1\perp}+\mathrm{t_2}\slsh{p}_{2\perp}+\mathrm{t_1}\mathrm{t_2}\slsh{\hat{F}}w }{\mathrm{t_1}\mathrm{t_2}\mathrm{t_3}-\mathrm{t_1}-\mathrm{t_2}-\mathrm{t_3}},
\end{align}
and $s\equiv s_1+s_2+s_3$. As we mention before, since the integration over the all momenta has been performed, the above result is general and does not depend on the integration order over the coordinate space (see Eq.~(\ref{eq.fase1}) and discussion bellow). Recall that it  corresponds to the diagram I and the analogous expressions for diagram II can be obtained by charge conjugation, as shown in Eq.~(\ref{eq.chargueconj}).

It is worth to remark that within the dimensional regularization scheme, there are not divergent terms in the vacuum case as shown in Appendix ~\ref{ap.details}. Therefore, Eq.~(\ref{eq.MF-VEdiagA4}) and the subsequent expressions has not divergent terms.

Beside the interaction particle modifications introduced by the magnetic field, an important contribution of the present work is the development of a procedure that allow us to carry out momentum integration preserving the spinorial trace and without any kinematic approximation. This approach give rise to a more manageable result since its is written in terms of integrals that can be worked out by applying standard procedures once certain approximations are performed. Note that our procedure allow us to obtain an equation which is valid for an arbitrary magnetic field strength. The methodology presented here is novel and so far, to our knowledge, has never been reported in the literature.

Although Eq.~(\ref{eq.MF-VEdiagA4}) is an exact result, the remaining integrals cannot be calculated analytically because of its intricate form. To gain some insight about the magnetic field effect on the analytical structure of the effective vertex, in the next section we shall explore three different magnetic field strength regions.

\section{Vertex behavior on different field strength regions} 
\label{sec.aproximations}

\noindent Taking into account that in Eq.~(\ref{eq.MF-VEdiagA4}) the physical scales are the VB momenta $p_{i}$, the fermion mass $m$ and the magnetic strength interaction $|qB|$, different approximations can be done depending on the hierarchy of scales among these quantities. In what follows, we verify the zero magnetic field limit and analyze the most recurred regions addressed in the literature, the strong and weak magnetic field strengths.

\subsection{Zero magnetic field limit}
\label{subsec.aproxZero}

\noindent Let us start by verifying that the vacuum case can be obtained from Eq.~(\ref{eq.MF-VEdiagA4}) by taking the limit $|qB|\longrightarrow0$. In this limit, the behavior of the different functions that appears in the above expressions go as follows
\begin{align*}
	\cos(qBs_j)\approx1, \ \ \ \tan(qBs_j)\approx qBs_j \ \ \ \text{and} \ \ 
 \ e^{(j)}\approx\mathbf{I}.
\end{align*}
Thus, once we replace these behaviors in  Eq.~(\ref{eq.MF-VEdiagA4}), and after some simple manipulations, it reduces to
\begin{equation}
	\begin{split}
		\lim_{|qB|\rightarrow0} i{\mathcal{M}_{{qB}}^{\mu\nu}}_{(I)}(p_1,p_2)=&ih g^2\left(-i\right)^{D/2+1}\frac{1}{2^D\pi^{D/2}}\\
		&\times\int dv_1\ dv_2 \ dv_3 \ \delta(1-v_1-v_2-v_3)\int_0^\infty ds \ s^{2-D/2} \ e^{i s \left(v_1v_3p_{1}^2+v_2v_3p_{2}^2+v_1v_2w^2-m^2\right)}\\
		&\times\Bigg\{\textit{Tr}\left[\gamma^\mu\left(m-{v_3\slsh{p}_1-v_2\slsh{w}}\right)\left(m+{v_1\slsh{w}+v_3\slsh{p}_2}\right)\gamma^\nu\left(m+{v_1\slsh{p}_1-v_2\slsh{p}_2}\right)\right]\\
		&\hspace{15mm}+\frac{im}{2s}\bigg(-D_\parallel\textit{Tr}\left[\gamma^\mu\gamma^\nu\right]+2\textit{Tr}\left[\gamma^\mu_\parallel\gamma^\nu\right]+2\textit{Tr}\left[\gamma^\mu\gamma^\nu_\parallel\right]\bigg)\\
		&\hspace{29mm}+\frac{im}{2s}\bigg(D_\perp\textit{Tr}\left[\gamma^\mu\gamma^\nu\right]-2D_\perp\textit{Tr}\left[\gamma^\mu\gamma^\nu\right]+2\textit{Tr}\left[\gamma^\mu\gamma^\nu_\perp\right]+2\textit{Tr}\left[\gamma^\mu_\perp\gamma^\nu\right]\bigg)\Bigg\},
	\end{split}
	\label{eq.MF-VEdiagAqB->0-1}
\end{equation}
where we used $s\equiv s_1+s_2+s_3$ and made the change of variable $s_j=s v_j$, with $s\in[0,\infty)$ and $v_j\in[0,1]$ in such a way the relation $v_1+v_2+v_3=1$ is fulfilled.

We left Eq.~(\ref{eq.MF-VEdiagAqB->0-1}) without any further simplifications in order to keep track on how the multiple trace terms in Eq.~(\ref{eq.MF-VEdiagA4}) contribute to the effective vertex in vacuum. In particular, the last two lines in the above equation simplify to
\begin{equation}
        {N_{{\text{vac.}}}^{\mu\nu}}_{(I)}=\frac{im}{2s}\bigg(\left(4-D\right)\textit{Tr}\left[\gamma^\mu\gamma^\nu\right]\bigg).
	\label{eq.MF-VEdiagAqB->0divergent}
\end{equation}
This term highlights the importance to work within the scheme of dimensional regularization~\cite{Gegelia}, since it shows up a logarithmic divergence, due the extra $\frac{1}{s}$ factor, that must be treated carefully\footnote{Recall that, in the Schwinger proper time formalism, the UV divergent behavior is translated to the $s\longrightarrow0$ region.}. The integration over $s$ in Eq.~(\ref{eq.MF-VEdiagAqB->0-1}) can be performed by standard methods and, in this particular case, the effective vertex does not show up any divergent behavior in the limit $D\longrightarrow4$ as shown explicitly in Appendix~\ref{ap.details}, so, it does not need any further treatment. After the remaining calculations are performed, we arrive at the final expression given by Eq.~(\ref{eq.sinB-EstructuraVerticeFinal}).

As we mention before, in Eq.~(\ref{eq.MF-VEdiagAqB->0-1}) we can easily track which factors in~Eq.~(\ref{eq.MF-VEdiagA4}) contribute to the divergent term in Eq.~(\ref{eq.MF-VEdiagAqB->0divergent}), these are
\begin{equation}
	\begin{split}
		{N_{{qB}}^{\mu\nu}}_{(I)}=&\dfrac{im}{2s}\bigg(-D_\parallel\textit{Tr}\left[\gamma^\mu e^{(1)}e^{(2)}\gamma^\nu e^{(3)}\right]+2\textit{Tr}\left[\gamma^\mu_\parallel e^{(1)}e^{(2)}\gamma^\nu e^{(3)}\right]+2\textit{Tr}\left[\gamma^\mu e^{(1)}e^{(2)}\gamma^\nu_\parallel e^{(3)}\right]\bigg)\\
		&+\frac{im \ qB }{2\left(\mathrm{t_1}\mathrm{t_2}\mathrm{t_3}-\mathrm{t_1}-\mathrm{t_2}-\mathrm{t_3}\right)}\bigg(-D_\perp\textit{Tr}\left[\gamma^\mu\gamma^\nu e^{(3)}\right]+D_\perp\textit{Tr}\left[\gamma^\mu\left(e^{(1)}+e^{(2)}\right)\gamma^\nu\right]\\
		&\hspace{76mm}-2\textit{Tr}\left[\gamma^\mu e^{(1)}\gamma^\nu_\perp\right]-2\textit{Tr}\left[\gamma^\mu_\perp e^{(2)}\gamma^\nu\right]\bigg).
	\end{split}
	\label{eq.MF-div}
\end{equation}
 From this equation, we can observe how the magnetic field splits the divergent terms into parallel and perpendicular structures. Note that the magnetic field does not lead to new divergences, {\it i.e.}, all these exclusively come from the vacuum, as pointed out in the Schwinger's seminal paper~\cite{SCWING}.

In what follows, this analysis shall give us an insight on how to deal with the divergent terms in two important regions widely used in the literature: the strong and weak magnetic field strength regions.

\subsection{Strong magnetic field limit}
\label{subsec.aproxStrong}

 \noindent The strong magnetic field regime is usually obtained by imposing that the magnetic field strength is way larger than the fermion mass, \textit{i.e.}, $|qB|\gg m^2$. Nevertheless, there is another energy scale that plays an important r\^{o}le in the effective vertex structure, the VB perpendicular momentum $p_{i\perp}$, which can be seen in Eq.~(\ref{eq.MF-VEdiagA4}) where there is an exponential factor that involves a combination of the perpendicular momentum and the magnetic field, given by
\begin{equation}
    e^{-\frac{i}{qB}\frac{\mathrm{t_1}\mathrm{t_3}p_{1\perp}^2+\mathrm{t_2}\mathrm{t_3}p_{2\perp}^2+\mathrm{t_1}\mathrm{t_2}w_{\perp}^2+2\mathrm{t_1}\mathrm{t_2}\mathrm{t_3}p_2\hat{F}p_1}{\mathrm{t_1}\mathrm{t_2}\mathrm{t_3}-\mathrm{t_1}-\mathrm{t_2}-\mathrm{t_3}}},
\label{exptsai}    
\end{equation}
so, any approximation performed on this term should be done with care~\cite{TSAI.1,TSAI.2}. In the particular case when the magnetic field strength is the highest energy scale, {\it i.e.},
\begin{equation*}
    m^2, |p_{i\perp}|^2 \ \ll \ |qB|
\end{equation*}
the exponential in Eq.~(\ref{exptsai}), can be safely approximated to one.

Next, to study the strong magnetic field limit of Eq.~(\ref{eq.MF-VEdiagA4}), following Ref.~\cite{Gusynin}, let us transfer it to Euclidean space by performing the replacements $p^0\longrightarrow ip_{4}$ with $p$ any momentum, $\gamma^0\longrightarrow i\gamma_{4}$, $\gamma_i\equiv\gamma^i$ with $i=1,2,3$ and $s_j\longrightarrow-is_j$. Thus, once these substitutions are done, the strong magnetic field behavior of the effective vertex reads
\begin{equation}
	\begin{split}
		i\mathcal{M}_{E\hspace{1mm}\mu\nu}^{qB\hspace{0.5mm}(I)}(p_1,p_2)=&-ih g^2|qB|^{D_\perp/2}\dfrac{\left(-1\right)^{D_\perp}}{\pi^{D/2}4^{D_\perp-1}2^{D_\parallel-1}}\\
        &\times\int_{1/\Lambda^2}^{\infty} \ s^{2-D_\parallel/2} ds\int dv_1\ dv_2 \ dv_3 \ \delta(1-v_1-v_2-v_3) \ e^{-s \left(v_1v_3p_{1\parallel}^2+v_2v_3p_{2\parallel}^2+v_1v_2w^2_{\parallel}+m^2\right)}\\
		&\times\Bigg\{\textit{Tr}\left[\gamma_{\parallel\mu}\left(m+{v_3\slsh{p}_{1\parallel}+v_2\slsh{w}_{\parallel}}\right)\left(m-{v_1\slsh{w}_{\parallel}-v_3\slsh{p}_{2\parallel}}\right)\gamma_{\parallel\nu}\left(m-{v_1\slsh{p}_{1\parallel}+v_2\slsh{p}_{2\parallel}}\right)\Delta_+\right]\\
        &\hspace{86mm}-\frac{m}{2s}\Bigg((4-D_\parallel)\textit{Tr}\left[\gamma_{\parallel\mu}\gamma_{\parallel\nu}\Delta_+\right]\Bigg)\Bigg\},
	\end{split}
	\label{eq.MF-VEdiagA4-StrongField2}
\end{equation}
where the subindex $E$ refers to the Euclidean space with $p^2\equiv p_4^2+\vec{p}^{ \, 2}$. $\Lambda$ is an ultraviolet cutt-off from which the magnetic field strength is the highest energy scale. $\Delta_+$ is one of the spin projectors defined as
\begin{equation}
    \Delta_\pm\equiv\dfrac{1}{2}\left(\mathbf{I}\pm i\hspace{1mm}\text{sign}(qB)\gamma_1\gamma_2\right),
    \label{eq.proyectores}
\end{equation}
which synthesize the behavior of the spin-magnetic field interaction factor
\begin{equation}
    e^{qB s_j\Sigma_3}=e^{|qB|s_j}\Delta_+ + e^{-|qB|s_j}\Delta_-,
    \label{proyectores}
\end{equation}
in the strong magnetic field limit, in Euclidean space.

Notice that in Eq.~(\ref{eq.MF-VEdiagA4-StrongField2}) only appears $\Delta_+$ which indicates that the fermions with spin along the magnetic field direction are the only ones who contributes to the process. As a consequence, the result is expressed only with parallel structures. The absence of perpendicular structures indicates that the fermions are constrained to the LLL, this phenomenology is well known as dimensional reduction~\cite{Gusynin}. An important aspect to highlight is that, in this limit, the result has a linear dependence with the magnetic field strength once the limit $D\longrightarrow4$ is taken.

Nevertheless, there is a different kinematic regime that can be studied when the VB perpendicular momenta is no longer a soft energy scale. In this region, the fermions acquire perpendicular dynamics since the hierarchy energy scale becomes
\begin{equation*}
    m^2\ll|qB|\lesssim|p_{i\perp}|^2,
\end{equation*}
and allow them to transit to higher Landau levels. However, in this kinematic region, these configurations are exponentially suppressed by the behavior of Eq.~(\ref{exptsai}) in Euclidean space.

The remaining integrals in Eq.~(\ref{eq.MF-VEdiagA4-StrongField2}) are quite similar to the vacuum case (see Appendix~\ref{ap.details}), thus, the effective vertex structure obtained for strong magnetic field limit would have the form
\begin{equation}
    \mathcal{M}_{E\hspace{1mm}\mu\nu}^{qB\hspace{0.5mm}}=\Tilde{A}_1\delta_{\parallel\mu\nu}
    +\Tilde{A}_2p_{1\parallel\mu} \ p_{2\parallel\nu}
    +\Tilde{A}_3p_{1\parallel\nu} \ p_{2\parallel\mu}
    +\Tilde{A}_4p_{1\parallel\mu} \ p_{1\parallel\nu}
    +\Tilde{A}_5p_{2\parallel\mu} \ p_{2\parallel\nu},
    \label{eq.tensorstronglimit}
\end{equation}
resembling the tensor decomposition
discussed in Sec.~\ref{subsec.tensorvac}. By projecting Eq.~(\ref{eq.tensorstronglimit}) onto the orthogonal basis (see Eq.~(\ref{eq.coefa1})), we can observe that the only contribution to the amplitude comes from VB in the $L^{*}$ polarization state.

As a final note, let us remark that there are not divergences in Eq.~(\ref{eq.MF-VEdiagA4-StrongField2}) since the region $s\longrightarrow0$ is excluded in the strong field limit.

\subsection{Weak magnetic field approximations}
\label{subsec.aproxWeak}

 \noindent By considering the magnetic field strength as the lowest energy scale, one locates at the weak magnetic field regime, {\it i.e.,}
\begin{equation*}
    |qB| \ \ll \ m^2, \  |p_{i\perp}|^2.
\end{equation*}
Once again, the exponential factor in Eq.~(\ref{exptsai}) have to be carefully treated since it is a function of three energy scales and whose behaviour depends on the hierarchy between $m$ and $p_{i\perp}$. In what follows, we discuss two different approximations depending on the VB kinematics~\cite{Piccinelli,Jaber}, however, we will not write explicitly the full analytical expressions for the effective vertex since they are extensive. A more exhaustive analyses can be found in Appendix~\ref{ap.details}.

Let us start by taking the hierarchy where the fermion mass is the highest energy scale, it is
\begin{equation*}
     |qB|\ll m^2 \ \ \text{and} \ \ |p_{i\perp}|^2\lesssim m^2.
\end{equation*}
Under this physical conditions, each term in Eq.~(\ref{eq.MF-VEdiagA4}) can be expanded in a Taylor series on $qB$, then, the effective vertex can be expressed as
\begin{equation}
    \mathcal{M}_{qB}^{\mu\nu}(p_1,p_2)=\mathcal{M}_{\text{vac.}}^{\mu\nu}(p_1,p_2)+{\Tilde{\mathcal{M}}_{qB}}^{\mu\nu}(p_1,p_2),
    \label{weaklow}
\end{equation}
where the first term correspond to the vacuum (see Eq.~(\ref{eq.MF-VEdiagAqB->0-1})) and the second is a polynomial function on $qB$ that contains all the magnetic field effects on the process. Due to Furry's theorem, the polynomial function only contains even powers of $qB$ since the odd powers cancel out once the two Feynman diagrams, shown in Fig.\ref{fig.diagramasChidosCampo}, are considered. In this approximation, named weak field with low perpendicular momentum approximation, the vacuum contribution is isolated, so, the treatment of the divergences is not modified by the magnetic field.

Next, if we take the VB perpendicular momentum as the highest energy scale, {\it i.e.},
\begin{equation*}
    |qB|\ll m^2<|p_{i\perp}|^2,
\end{equation*}
we cannot a perform a Taylor series expansion in the exponential term in Eq.~(\ref{exptsai}), as was done previously, since there are factors of $p_{i\perp}$ that make the exponential argument larger than one. However, a power expansion on $qB$ in the trigonometric functions in Eq.~(\ref{eq.MF-VEdiagA4}) is valid.

In this physical regime, called weak field with high perpendicular momentum, a separation similar to Eq.~(\ref{weaklow}) can not be done since the vacuum contributions are mixed with the magnetic ones. The $qB$ dependence is not fully contained in the polynomial function, so, the UV-divergences  can not be separated as in the previous case.

The weak field with low perpendicular momentum expansion in Eq.~(\ref{weaklow}) can be obtained by a different approach, commonly used in literature, where the fermion propagators dressed with the magnetic field effects are expanded in a power series on $qB$ at the beginning of the calculation~\cite{TaiwanDebil,AyalaDebilGluon,AyalaDebilPionMass}. As shown in Ref.~\cite{Jaber}, both approaches lead to the same results, however, in the way we performed the approximation it became explicit that the results are restricted to one the low perpendicular momentum regime. In this sense, the methodology presented in this work is more general because allows us to analyze different magnetic field regions and kinematics regimes from an analytical expression valid for an arbitrary magnetic field strength. Within the weak field approximation at low perpendicular momentum, in the next section we shall study the effect of the magnetic field on the cross section.

\section{Insight of the magnetic field effect on the interaction}
\label{sec.results}

\noindent In order to get some insight of the magnetic field effect on the VB's-scalar interaction, let us focus on the scalar bosons production through VB fusion, $V^\mu+V^\nu\longrightarrow\phi$, in the weak field with low perpendicular momentum approximation, where the observable of interest is the cross section. To isolate the magnetic field effect, we shall compare the quantities dressed with the magnetic field  with those obtained at zero magnetic field (by taking its ratio).

With this goal in mind, we shall use the procedure discussed in Sec.\ref{subsec.aproxWeak} and Sec. \ref{subsec.apenIntegration.weak} to obtain the effective vertex behavior at weak magnetic field in the low perpendicular momentum regime. Then, the obtained result is used to get the form factors $a_1$, $a_2$ and $a_4$, up to order $\mathcal{O}\left((qB)^2\right)$, by projecting the amplitude with each orthogonal tensorial structure, as shown in Eq.~(\ref{eq.coefa1}). Finally, the cross section is computed by using Eq.~(\ref{eq.sinB-crosssectionFinal}).

Next, for the sake of simplicity, let us consider a frontal VB collision fully contained in the perpendicular plane, to the magnetic field direction. As a result, the scalar bosons are produced along the ``collision line''. In this collision, the kinematic configuration for VB momenta is given by: $p_{1,z}=p_{2,z}=0$ and the angle between the VB in the perpendicular plane, $\theta$, is equal to $\pi$. Moreover, in order to assure that the produced scalar particles are real, the values of $p_{1\perp}$ and $p_{2\perp}$ must satisfy the on-shell condition 
\begin{equation}
    p_1\cdot p_2=\dfrac{m_\phi^2}{2},
    \label{eq.onshellmass}
\end{equation}
\noindent which is a consequence of the energy-momentum conservation, $p_1^\mu+p_2^\mu=w^\mu$.

\begin{figure}[H]
	\centering
	\subfigure[Magnetic field effect on the form factor $a_1$.]{\includegraphics[width=8cm]{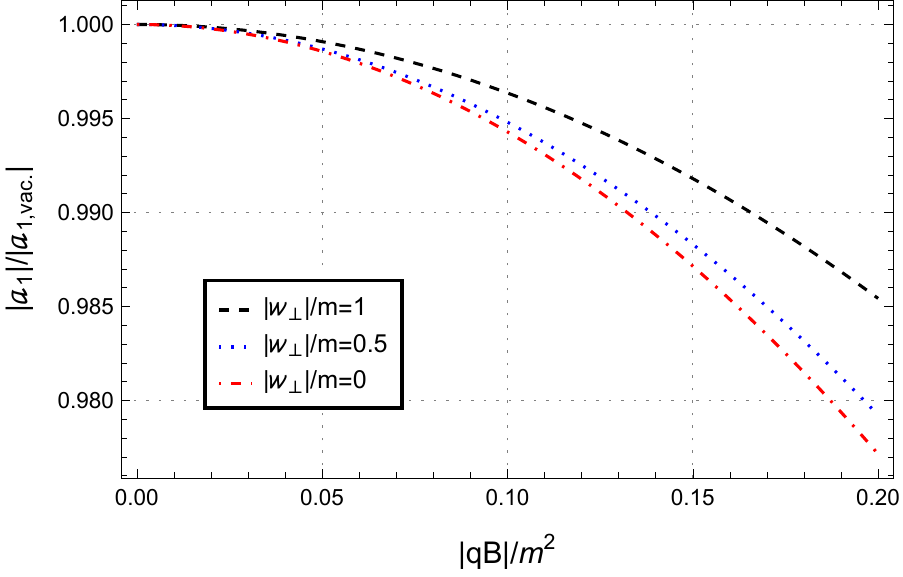} }%
	\qquad
	\subfigure[Magnetic field effect on the form factor $a_2$.]{\includegraphics[width=8cm]{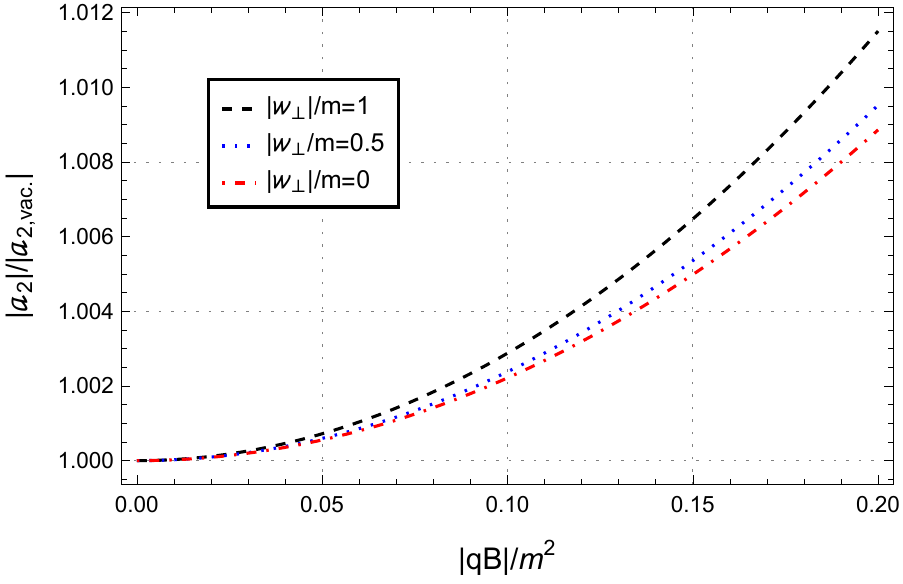} }%
	\caption{Form factor behavior, for a VB head to head collision fully contained in the perpendicular plane within the weak field approximation at low perpendicular momentum, as a function of the magnetic field strength $|qB|$ for different scalar particle perpendicular momentum values $|w_\perp|$. Taking $p_{i,z}=0$, $\theta=\pi$ and $m/m_\phi=170/125$.}
	\label{fig.coefweakfield}
\end{figure}

\begin{figure}[H]
	\centering
	\includegraphics[width=9cm]{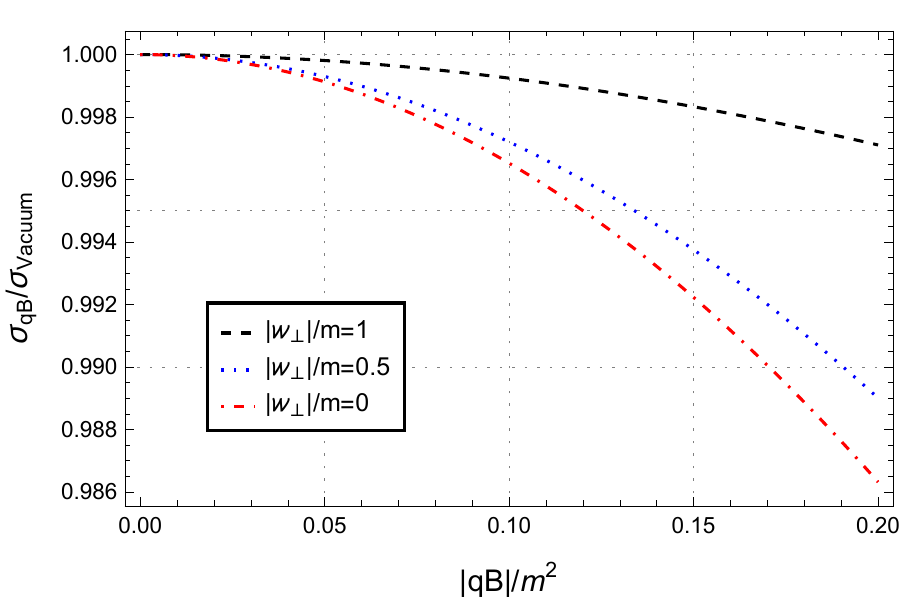}
	\caption{Cross section behavior, for a VB head to head collision fully contained in the perpendicular plane within the weak field approximation at low perpendicular momentum, as a function of the magnetic field strength $|qB|$ for different scalar particle perpendicular momentum values $|w_\perp|$. Taking $p_{i,z}=0$, $\theta=\pi$ and $m/m_\phi=170/125$.}
	\label{fig.secefiweakfield}
\end{figure}

In Figs.~\ref{fig.coefweakfield}-\ref{fig.secefiweakfield}, it is shown the behavior of the form factors $a_1$, $a_2$ and the cross section as a function of the magnetic field strength for different values of the scalar perpendicular momentum $|w_\perp|$, taking $m/m_\phi=170/125$. Note that the red curves correspond to the case in which the scalars are produced at rest.

Fig.~\ref{fig.coefweakfield} shows the effect of the magnetic field on the form factors. In Fig.~\ref{fig.coefweakfield}(a), the form factor $a_1$ decreases as the magnetic field strength grows, meanwhile, in Fig.~\ref{fig.coefweakfield}(b), the form factor $a_2$ increases with the magnetic field strength. This indicates that the contribution due VB with $L$-state polarizations is reduced by the magnetic field, meanwhile $L^*$-state contribution is enhanced. Furthermore, both form factors are enhanced as the scalar perpendicular momentum increases. Note that, in Fig.~\ref{fig.coefweakfield}, the form factor $a_4$ is not plotted since it identically vanishes, indicating that the collision of VB in different polarization states (one in the $L$-state and the other in the $L^*$-state) does not contribute to the amplitude.

The effects of the external magnetic field on the cross section are shown in Fig.~\ref{fig.secefiweakfield}. This figure shows that the cross section decreases as the magnetic field strength grows. This indicates that the production rate is reduced by the presence of the magnetic field. Also, as in the previous figure, the cross section increases as the scalar perpendicular momentum grows.

Taking a closely look at Fig.~\ref{fig.coefweakfield} and Fig.~\ref{fig.secefiweakfield}, one can notice that the main magnetic field effect on the cross section comes from the form factor $a_1$, indicating that the polarization $L$-states are more sensitive to the magnetic field presence.

As previously mentioned, the results shown in this section just give us an idea about the magnetic field effect on the VB's-scalar interaction for a particular process. A more exhaustive analysis is a work in progress and will be reported elsewhere.

\section{Conclusions}
\label{sec.concl}

\noindent This paper presents a novel methodology to compute the one-loop amplitude of a scalar particle production through massless vector bosons fusion, $V^\mu+V^\nu\longrightarrow\phi$, when an homogeneous magnetic field, with arbitrary strength, is present. This methodology can be extended to compute an arbitrary particle process, with any number of internal charged or external neutral particles, in presence of an external magnetic field.

By using the polarization states of the VB, relevant in the physical scenario we were interested in, together with general physical considerations that must be satisfied in this particles interaction, we obtained a general tensor structure for the amplitude process, written in terms of orthogonal structures. This last point made smooth the calculation of physical observables, reducing it to the explicit computation of the amplitude.

Within a QED-based model, to study the interaction strength between a scalar particle and two VB in a magnetic background, we focused our attention on leading order contributions, corresponding to triangular Feynman diagrams with a charged fermion loop. We incorporated the magnetic field into the process by employing the Schwinger's proper time fermion propagators. These contributions were computed for an arbitrary magnetic field strength by carrying out the phase-space integration in $D$-dimensions without performing the spin trace explicitly. The final result was expressed in terms of Schwinger's proper time, which has a simple and compact form.

Along with the amplitude computation, we maintain a clear treatment of the Schwinger's phases and the factors that arise from them, gaining a deeper understanding of its r\^ole on the particles interaction.

By taking the zero magnetic field limit, the expressions obtained with our methodology match the results reported in the literature for the vacuum case (in the Higgs physics context), including the treatment of the divergent terms within the dimensional regularization scheme.

An advantage of having an exact result, expressed in terms of Schwinger's proper time instead of an infinite sum over Landau levels, is that allows us to study different limiting cases of the magnetic field strength without any restriction on the VB kinematics which is in contrast with a widely used approach followed in literature where any approximation, done at the beginning of the computation, restricts the study scope to only one kinematic and field strength region.

Our result led us aboard the r\^ole played by the kinematic of the particles involved in the interaction, revealing a broader energy scale hierarchy compared with the usual one. Then, a wider number of physical situations emerges: strong and weak field regions at low and high VB perpendicular momentum.

We found that, in the strong magnetic field limit, only parallel tensor structures are present in the amplitude and the perpendicular ones are exponentially suppressed. Furthermore, in this limit, the UV-divergences are absent since the region $s\longrightarrow0$ is excluded, by a physical energy scale $\Lambda$ from which this limit is valid.

In the weak field region at low VB perpendicular momentum, we were able to  separate the vacuum and the magnetic field contributions. This latter is divergent free, so, the UV-behavior is completely encode in the vacuum one. This separation can not be performed in the high VB perpendicular momentum approximation since there is an exponential factor that mixed the vacuum and magnetic field contributions.

By considering a frontal VB collision fully contained in the perpendicular plane, we analyzed the magnetic field effect on the scalar boson production process through VB fusion within the weak field approximation at low perpendicular momentum. Our findings indicates that the cross section is inhibited by the magnetic field presence, for a fixed scalar boson perpendicular momentum, and increases as scalar perpendicular momentum grows. This behavior is mainly inherited from the form factor $a_1$, highlighting that the polarization $L$-states are more sensitive to the magnetic field presence.

In conclusion, the methodology presented in this work simplifies the calculation of the scalar boson production amplitude within a magnetic background and could be extended for a more general particle process, with internal charged particles, in presence of an homogeneous magnetic field with an arbitrary strength, simplifying its calculation and extending the analysis scope to a wider range of particle kinematic regions.

\appendix

\section{Polarization vectors}
\label{ap.polarizacion}

\noindent The vectors that build the orthogonal basis, presented in Sec.~\ref{subsec.tensormag}, are well know in the literature (see for example~\cite{Kuznetsov}) and was introduced to describe photons in presence of an external magnetic field. In this appendix, we will discuss the physical aspects of the orthogonal basis and the VB polarization vectors.

\subsection{Sum over polarizations}

\noindent In general, the invariant amplitude can be written as
\begin{equation}
    \mathcal{M}\equiv\mathcal{M}^{\mu\nu}\epsilon_\mu(p_1,\lambda_1)\epsilon_\nu(p_2,\lambda_2),
\end{equation}
where $\epsilon_\mu(p,\lambda)$ are the polarization vectors of the VB. Then, in the particular case when the VB are on-shell, the summation over polarizations of the square amplitude in Eqs.~(\ref{eq.sinB-diferentialcrosssection}, \ref{eq.sinB-diferentialdecayrate}) is given by
\begin{equation}
    \sum_{\text{spin}}|\mathcal{M}|^2=\sum_{\lambda_1=\pm1}\sum_{\lambda_2=\pm1}\mathcal{M}^{\mu\nu}\mathcal{M}^{*\alpha\beta}\epsilon_\mu(p_1,\lambda_1)\epsilon_\nu(p_2,\lambda_2){\epsilon_\alpha^*}(p_1,\lambda_1){\epsilon_\beta^*}(p_2,\lambda_2),
	\label{eq.sinB-SumaEspinesGeneral}
\end{equation}
where, for the sake of simplicity, the sum of the physical polarization has been written in the linear polarization basis.

By using the widely known property
\begin{equation}
	\begin{split}
	    \sum_{\lambda=\pm1}\epsilon_\mu(p,\lambda)\epsilon_\alpha^{*}(p,\lambda)&=-g_{\mu\alpha}+\dfrac{p_\mu a_\alpha+p_\alpha a_\mu}{p\cdot a}+a^2\dfrac{p_\mu p_\alpha}{\left(p\cdot a\right)^2}\\
     &\equiv \mathcal{P}^{\mu\nu}\left(p,a\right),
	\end{split}
	\label{eq.sinB-SumaPlarizacionesGeneral}
\end{equation}
with $a^\mu$ an arbitrary auxiliary vector (which is $p^\mu$-independent), Eq.~(\ref{eq.sinB-SumaEspinesGeneral}) can be rewritten as
\begin{equation}
    \sum_{\lambda_1=\pm1}\sum_{\lambda_2=\pm1}|\mathcal{M}|^2=\mathcal{M}^{\mu\nu}\mathcal{P}_{\mu}^{\hspace{1.5mm}\alpha}\left(p_1,a\right)\mathcal{M}_{\alpha\beta}^{*}\mathcal{P}_{\nu}^{\hspace{1.5mm}\beta}\left(p_2,b\right).
	\label{eq.sinB-PromEspin}
\end{equation}
This result holds for any on-shell VB independently if there is or not an external magnetic field. With this in mind, the sum over physical polarizations of the square amplitude in presence of an external magnetic field describe by Eq.~(\ref{eq.MF.EsttensorOnShell}), gives
\begin{equation}
    \sum_{\lambda_1=\pm1}\sum_{\lambda_2=\pm1}|\mathcal{M}_{{qB}}|^2=|a_1^{++}|^2+|a_2^{++}|^2+|a_4^{+-}|^2,
	\label{eq.conB-PromEspin}
\end{equation}
where the on-shell conditions, $p_1^2=p_2^2=0$, were used.

Note that the final result is independent of the choice of the two auxiliary vectors $a^\mu$ and $b^\mu$.

\subsection{Polarization vectors}

\noindent The relation given in Eq.~(\ref{eq.sinB-SumaPlarizacionesGeneral}) has been written in the linear polarization basis for simplicity proposes, however, must be valid for any physical polarization vectors.

\begin{figure}[H]
	\centering
	\subfigure[Arbitrary case for Adler's polarization vectors.]{\includegraphics[width=7.5cm]{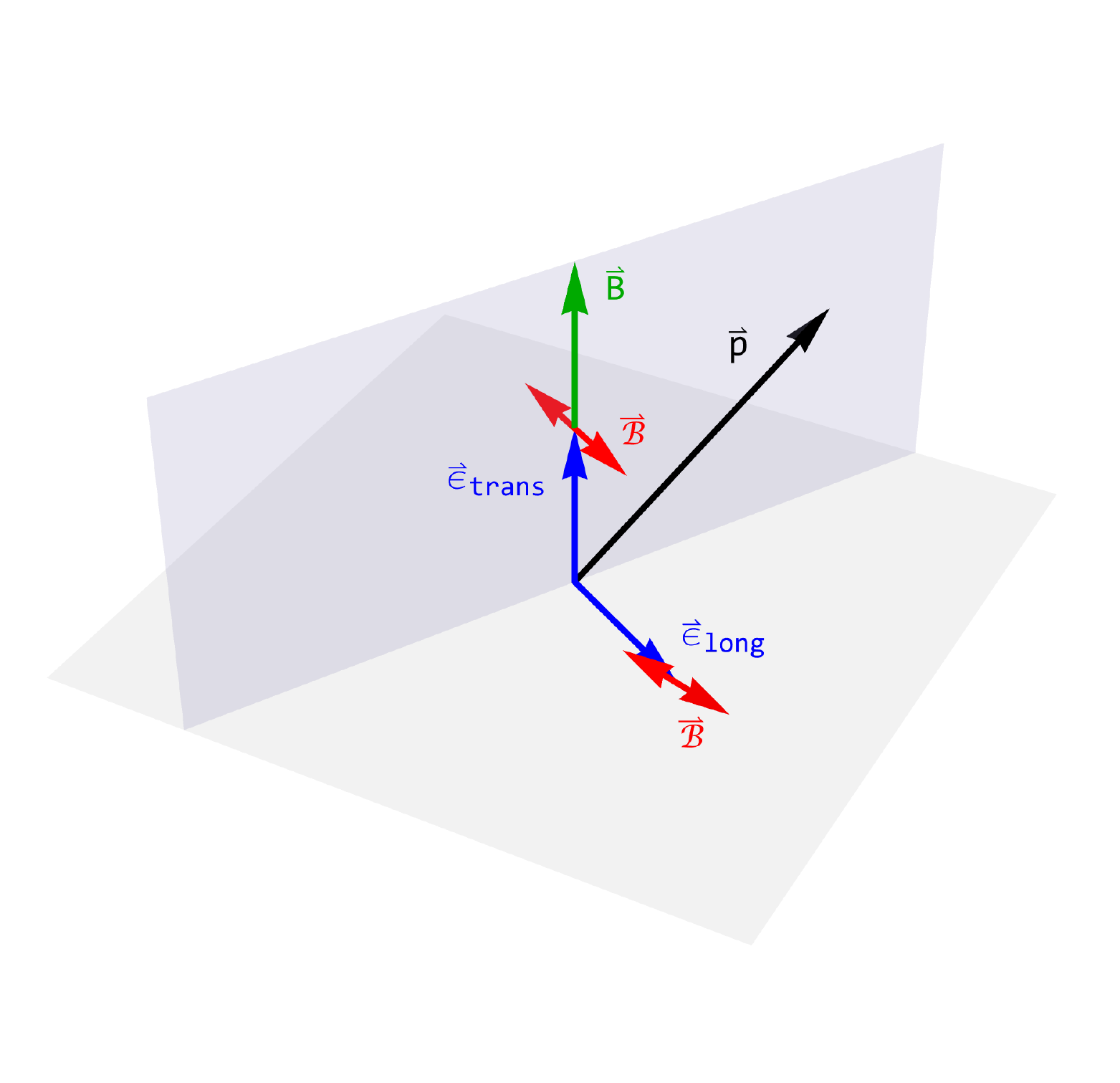} }%
	\qquad
	\subfigure[$\vec{p}\perp\vec{B}$ case for Adler's polarization vectors.]{\includegraphics[width=7.7cm]{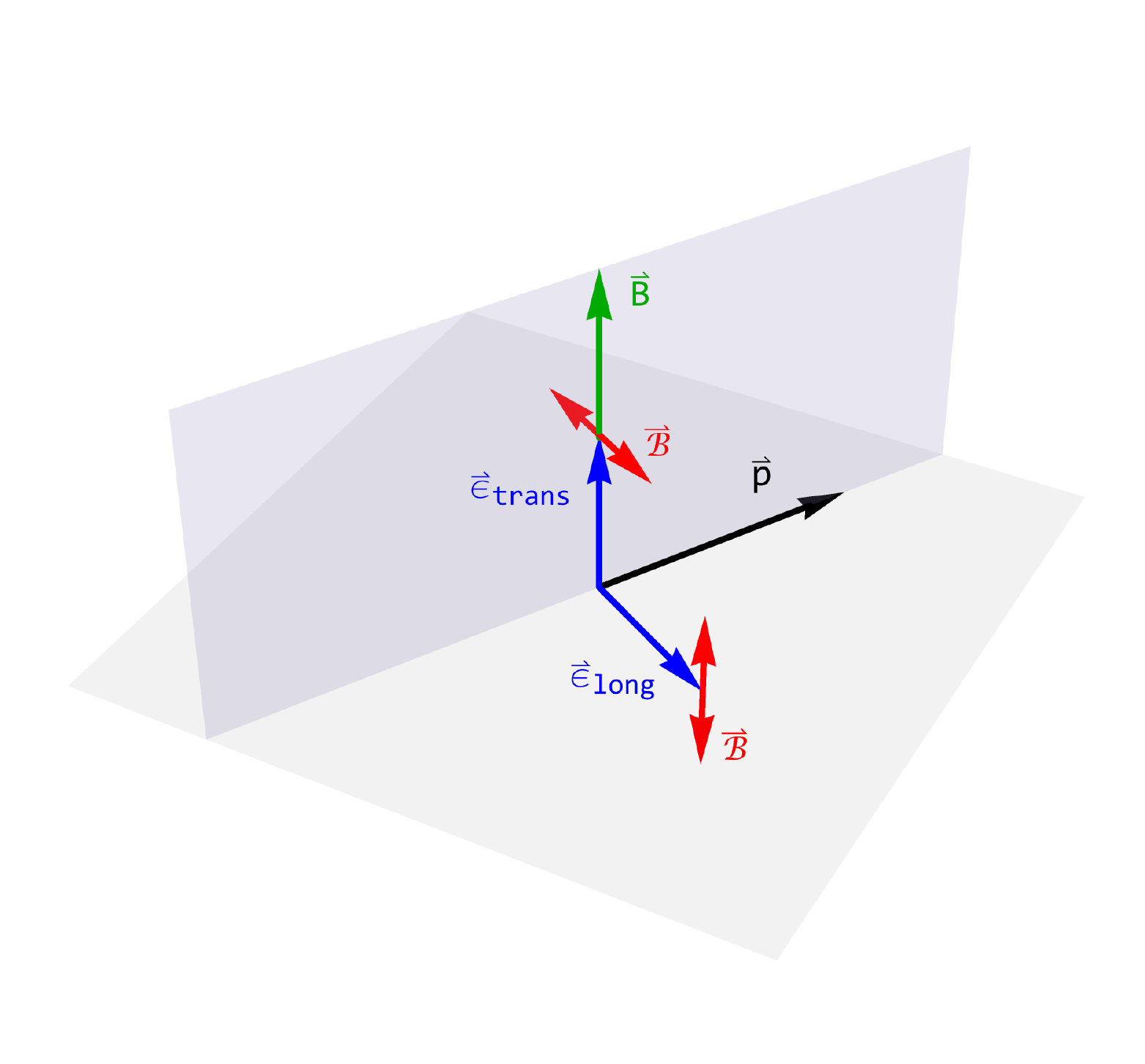} }%
 	\qquad
	\subfigure[Arbitrary case for the modified polarization vectors.]{\includegraphics[width=7.5cm]{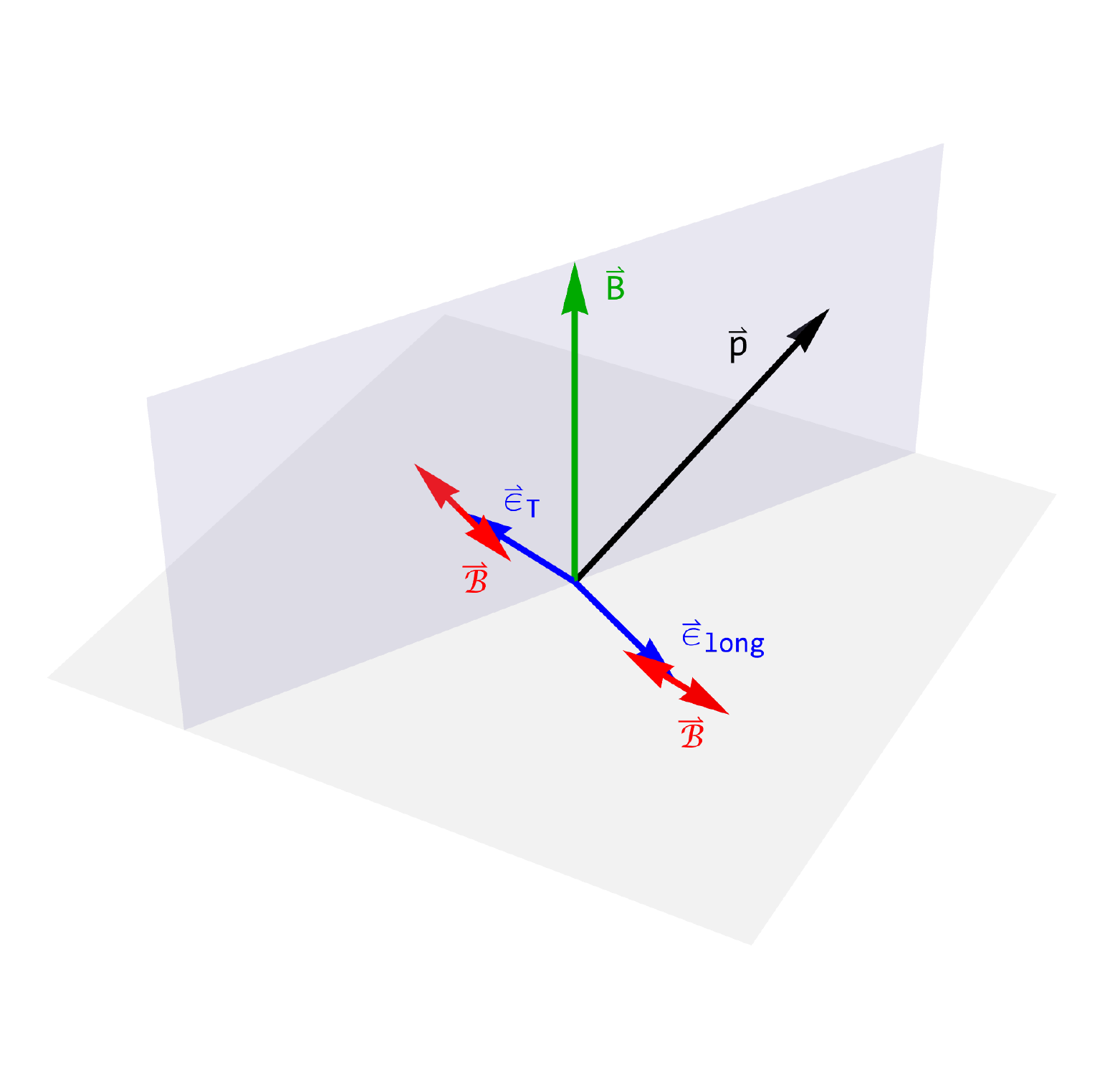} }%
	\caption{Polarization vectors for different physical situations. The following arrow color code is implemented: green for the external magnetic field $\vec{B}$, black for the VB momentum $\vec{p}$, blue for the polarization vectors $\vec{\epsilon}$ and red for its corresponding generalized magnetic-like field $\vec{\mathcal{B}}$. The generalized electric-like field $\vec{\mathcal{E}}$ goes along its corresponding polarization vector direction.}
	\label{fig.polvectors}
\end{figure}

In literature~\cite{Kuznetsov,Dittrich}, the so called polarizations vectors  associated with the basis introduced in Eq.~(\ref{setspolvects}), are
\begin{equation}
	\epsilon_{\text{long.}}^{\mu}\equiv\frac{L^{\mu}}{\sqrt{-p_\perp^2}} \ \ \ \text{and} \ \ \	\epsilon_{\text{trans.}}^{\mu}\equiv\frac{L^{*\mu}}{\sqrt{p_\parallel^2}},
    \label{eq.conB-polarizacionesRitus}
\end{equation}
where the notation refers to the ``longitudinal'' and ``transversal'' modes, according to Adler's decomposition~\cite{Adler}, for on-shell photons. This nomenclature indicates the position of the magnetic field vector of the photon wave with respect to the plane spanned by the photon momentum and the external magnetic field, as it is shown in Fig. \ref{fig.polvectors}a where the following arrow color code is implemented: green for the external magnetic field $\vec{B}$, black for the VB momentum $\vec{p}$, blue for the polarization vectors $\vec{\epsilon}$ and red for its corresponding generalized magnetic-like field $\vec{\mathcal{B}}$. As we are not working with specific VB, the modes refers to the orientations of $\vec{\mathcal{B}}$ of their associated wave.

It worth to note that the vectors in Eq.~(\ref{eq.conB-polarizacionesRitus}) do not fulfills Eq.~(\ref{eq.sinB-SumaPlarizacionesGeneral}) because they are not fully orthogonal to the VB momentum, in a 3 dimension sense, as shown in Fig. \ref{fig.polvectors}a. In this figure, meanwhile $\vec{\mathcal{B}}$ is orthogonal to $\vec{p}$ for both modes, the generalized electric-like field $\vec{\mathcal{E}}$ (that defines the polarization vector direction) is not orthogonal to the momentum for the ``transversal'' mode
\begin{equation}
    \vec{\epsilon}_{\text{trans.}}\cdot\vec{p}\neq0,
\end{equation}
 as it is also observed in Ref.~\cite{Hattori}, in this way, this mode does not describe a pure physical polarization. The situation changes when $\vec{p}$ is perpendicular to $\vec{B}$, as we can see in Fig. \ref{fig.polvectors}b. There, the basis becomes fully orthogonal and Eq.~(\ref{eq.sinB-SumaPlarizacionesGeneral}) is fulfilled.

Since the polarization vectors in Eq.~(\ref{eq.conB-polarizacionesRitus}), that form now on we take the liberty to call Adler's polarization vectors, do not correspond to the physical polarization ones in all cases, then, one has to construct an alternative polarization vector for the general case. This can be done by replacing $\vec{\epsilon}_{\text{trans.}}$ with
\begin{equation}
    \vec{\epsilon}_T=\dfrac{1}{|\vec{p}||\vec{L}|}\vec{L}\times\vec{p}.
\end{equation}
\noindent This new vector can be generalized to a four-vector as follows
\begin{equation}
    \epsilon^\mu_T=\dfrac{1}{E_{p}\sqrt{-p_\perp^2}}\epsilon^{\mu\nu\alpha\beta}n_\nu\hat{F}_{\alpha\sigma}p^\sigma p_\beta,
\end{equation}
whit $n^\nu\equiv(1,0,0,0)$. Since the modified polarization vectors are fully orthogonal, as we can observe in Fig. \ref{fig.polvectors}c, the relation (\ref{eq.sinB-SumaPlarizacionesGeneral}) is satisfied
\begin{equation}
\sum_{\lambda=\text{long.},T}\epsilon_\mu(p,\lambda)\epsilon_\alpha^{*}(p,\lambda)=-g_{\mu\alpha}+\dfrac{p_\mu n_\alpha+p_\alpha n_\mu}{p\cdot n}+n^2\dfrac{p_\mu p_\alpha}{\left(p\cdot n\right)^2}.
	\label{eq.sinB-SumaPlarizacionesGeneralMod}
\end{equation}

A different way to compute the Eq.~(\ref{eq.sinB-SumaEspinesGeneral}) is to project onto the polarization vectors before its sum is performed. In the magnetic field case, we write down the matrix element $\mathcal{M}_{qB}$ as the contraction of Eq.~(\ref{eq.MF.EsttensorOnShell}) and the modified polarization vectors associated to the ``longitudinal'' and $T$ modes. Then, by summing the different contributions, the result in Eq.~(\ref{eq.conB-PromEspin}) holds.

As a final remark, note that the polarization vectors based on the external  magnetic field breaks down if the VB momentum is aligned with the external magnetic field, $\vec{p}\parallel\vec{B}$. In that case, the tensor structure of the effective vertex in presence of a magnetic field obtained in Sec.~\ref{subsec.tensormag} is not valid. Nevertheless, the same analysis can be  done with a different set of vectors describing the VB polarizations.

\section{Schwinger's phase}
\label{ap.fase}

\noindent In this Appendix we show more details on the calculation of total Schwinger's phase for the Feynman diagrams shown in Fig. \ref{fig.diagramasChidosCampo}. With this in mind, let us start by writing the expression for the Schwinger's phase of a single propagator, given in  Eq.~(\ref{eq.conB-FaseSchwinger}) 
\begin{equation}
	\Omega(x',x'')=\exp\left(-iq\int_{x''}^{x'} A_\mu(x)dx^\mu\right).
 \label{schwingerphase}
\end{equation}

Now, in order to obtain the phase explicit form, we chose a four-potential $A^\mu(x)$, in such a way it generates an homogeneous magnetic field along $z$ direction, as follows
\begin{equation}
	A^\mu(x)=\dfrac{1}{2}x_\nu F^{\nu\mu}+\partial^\mu\chi(x),
	\label{eq.fase-NormaArbi}
\end{equation}
where $F^{\mu\nu}$ the field strength tensor with $F^{21}=-F^{12}=B$, the only non-vanishing components, and $\chi(x)$ an arbitrary function. Using Eq.~(\ref{eq.fase-NormaArbi}) in Eq.~(\ref{schwingerphase}), the Schwinger's phase associated with each fermionic propagator takes the form
\begin{equation}
	\Omega(x,y)=e^{i\frac{qB}{2}x_\mu \hat{F}^{\mu\nu}y_\nu}e^{-iq\left(\chi(x)-\chi(y)\right)},
	\label{eq.fase-FaseSchwRes1}
\end{equation}
where the integration over the coordinates has been performed along a straight line.

Finally, from the above result, it easy to see that the total Schwinger's phases for diagrams I and II, appearing in Eqs.~(\ref{eq.MF-VEdiagA}, \ref{eq.MF-VEdiagB}), read
\begin{equation}
    \begin{split}
    \Omega(x,y)\Omega(y,z)\Omega(z,x)&=e^{i\frac{qB}{2}\hat{F}^{\mu\nu}\left(x_\mu y_\nu+y_\mu z_\nu+z_\mu x_\nu\right)},\\
	\Omega(y,x)\Omega(x,z)\Omega(z,y)&=e^{-i\frac{qB}{2}\hat{F}^{\mu\nu}\left(x_\mu y_\nu+y_\mu z_\nu+z_\mu x_\nu\right)},
    \end{split}
 \label{eq.conB-FaseTotal}
\end{equation}
where
\begin{equation}
	\hat{F}^{\mu\nu}\left(x_\mu y_\nu+y_\mu z_\nu+z_\mu x_\nu\right)=-x_2y^1+x_1y^2-y_2z^1+y_1z^2-z_2x^1+z_1x^2.
	\label{eq.conB-xi}
\end{equation}

Note that the total phases are gauge invariant while the single Schwinger's phase on each propagator are not. This is a known fact for any charged closed loop in presence of an external electromagnetic field~\cite{Kuznetsov}. Another important point is that diagrams I and II have opposite phases, which give rise to interference terms that are not present in the vacuum case.

\section{Schwinger proper time integration} 
\label{ap.details}

\noindent In this Appendix we show details on how the integration over the Schwinger proper time is carried out for the different magnetic field strengths discussed in Sec.~\ref{sec.aproximations}. With this in mind, let us start with zero magnetic field limit.

\subsection{Zero magnetic field}
\label{subsec.apenIntegration.vacuum}

\noindent In the zero magnetic field limit, analyzed in Sec.\ref{subsec.aproxZero}, the effective vertex reads
\begin{equation}
	\begin{split}
		i{\mathcal{M}_{{\text{vac.}}}^{\mu\nu}}_{(I)}(p_1,p_2)=&ih g^2\left(-i\right)^{D/2+1}\frac{1}{2^D\pi^{D/2}}\\
		&\times\int_0^\infty ds \ s^{2-D/2}\int dv_1\ dv_2 \ dv_3 \ \delta(1-v_1-v_2-v_3) \ e^{i s \left(v_1v_3p_{1}^2+v_2v_3p_{2}^2+v_1v_2w^2-m^2\right)}\\
		&\times\Bigg\{\textit{Tr}\left[\gamma^\mu\left(m-{v_3\slsh{p}_1-v_2\slsh{w}}\right)\left(m+{v_1\slsh{w}+v_3\slsh{p}_2}\right)\gamma^\nu\left(m+{v_1\slsh{p}_1-v_2\slsh{p}_2}\right)\right]\\
		&\hspace{82mm}+\frac{im}{2s}\bigg(\left(4-D\right)\textit{Tr}\left[\gamma^\mu\gamma^\nu\right]\bigg)\Bigg\}.
	\end{split}
	\label{eq.MF-VEdiagAqB->0-1apen}
\end{equation}

In the above equation, the integration over $s$ can be performed straightforward by using the Gamma function integral representation, given by
\begin{equation}
	\int_{0}^{\infty}s^{n-\frac{D}{2}}e^{-i\Delta^2s}ds=\left(i\Delta^2\right)^{\frac{D}{2}-(n+1)}\Gamma\left(n+1-\dfrac{D}{2}\right),
	\label{eq.sinB-InteGamma1}
\end{equation}
\noindent obtaining
\begin{equation}
	\begin{split}
		i{\mathcal{M}_{{\text{vac.}}}^{\mu\nu}}_{(I)}(p_1,p_2)=&ih g^2\left(-i\right)^{D/2+1}\frac{1}{2^D\pi^{D/2}}\Gamma\left(3-\dfrac{D}{2}\right)\int dv_1\ dv_2 \ dv_3 \ \delta(1-v_1-v_2-v_3)\\
		&\times\Bigg\{\left(i\Delta^2\right)^{\frac{D}{2}-3}\textit{Tr}\left[\gamma^\mu\left(m-{v_3\slsh{p}_1-v_2\slsh{w}}\right)\left(m+{v_1\slsh{w}+v_3\slsh{p}_2}\right)\gamma^\nu\left(m+{v_1\slsh{p}_1-v_2\slsh{p}_2}\right)\right]\\
		&\hspace{70mm}+\frac{im}{4-D}\left(i\Delta^2\right)^{\frac{D}{2}-2}\bigg(\left(4-D\right)\textit{Tr}\left[\gamma^\mu\gamma^\nu\right]\bigg)\Bigg\},
	\end{split}
	\label{eq.MF-VEdiagAqB->0-1apen2}
\end{equation}
\noindent where $\varDelta_m ^2 \equiv m^2-v_1v_3p_{1}^2-v_2v_3p_{2}^2-v_1v_2w^2$.

In Eq.~(\ref{eq.MF-VEdiagAqB->0-1apen2}), we have not simplified the last term in order to emphasizes the importance to work within the dimensional regularization scheme. Once the ultraviolet behavior is regularized in the $s$ integration and the result do not have divergent terms, we can safely take $D\longrightarrow4$, getting
\begin{equation}
	\begin{split}
		i{\mathcal{M}_{{\text{vac.}}}^{\mu\nu}}_{(I)}(p_1,p_2)=&ih g^2\frac{1}{16\pi^{2}}\int_{0}^{1} dv_1\int_{0}^{1-v_1} dv_2\\
		&\times\Bigg\{\dfrac{\textit{Tr}\left[\gamma^\mu\left(m+(v_1-1)\slsh{p}_1-v_2\slsh{p}_2\right)\left(m+v_1\slsh{p}_1+(1-v_2)\slsh{p}_2\right)\gamma^\nu\left(m+v_1\slsh{p}_1-v_2\slsh{p}_2\right)\right]}{m^2-v_1(1-v_1-v_2)p_{1}^2-v_2(1-v_1-v_2)p_{2}^2-v_1v_2w^2}\\
        &\hspace{112mm}-m\textit{Tr}\left[\gamma^\mu\gamma^\nu\right]\Bigg\},
	\end{split}
	\label{eq.MF-VEdiagAqB->0-1apen3}
\end{equation}
\noindent where the integration over $v_3$ has also been done.

Adding both charge conjugate contributions and computing the trace, the two form factors in Eq.~(\ref{eq.sinB-EstructuraVerticeFinal}) can be identified as
\begin{align}
    A_1&=\dfrac{h g^2 m}{\pi^2}\int_{0}^{1} dv_1\int_{0}^{1-v_1} dv_2 \ \dfrac{4v_1v_2-1}{\tau-4v_1 v_2},\\
    A_2&=-\dfrac{2h g^2 m}{\pi^2m_\phi^2}\int_{0}^{1} dv_1\int_{0}^{1-v_1} dv_2 \ \dfrac{(2v_1-1)(2v_2-1)}{\tau-4v_1 v_2},
\end{align}
\noindent  where the external particles has been considered on-shell and we have introduced the notation $\tau\equiv4m^2/m_\phi^2$. This integrals has been done analytically by several authors~\cite{Marciano2012a,HiggsGamaGama,Spira} and can be expressed in terms of logarithmic, trigonometric or dilogarithmic functions ($\text{Li}_2$). In particular, in the latter form the coefficients are given by
\begin{align}
    \label{eq.A1vacio}
    A_1=&\dfrac{h g^2 m}{4\pi^2}\left[(\tau-1)\text{Li}_2\left(\dfrac{2}{1+\sqrt{1-\tau}}\right)+(\tau-1)\text{Li}_2\left(\dfrac{2}{1-\sqrt{1-\tau}}\right)-2\right],\\
    \label{eq.A2vacio}
    A_2=&-\dfrac{h g^2 m}{2\pi^2m_\phi^2}\left[8\sqrt{\tau-1}\cot^{-1}\left(\tau-1\right)+(\tau+1)\text{Li}_2\left(\dfrac{2}{1+\sqrt{1-\tau}}\right)+(\tau+1)\text{Li}_2\left(\dfrac{2}{1-\sqrt{1-\tau}}\right)-10\right],
\end{align}
\noindent which can be expressed as in Eqs.~(\ref{marciano})-(\ref{marciano3}), for the Higgs context. In this way, we have corroborated that our procedure reproduces the vacuum case results reported in the literature.

Even though, in the vacuum case there is not a privileged spatial direction, we still have the possibility to describe the VB polarization states with respect to an arbitrary spatial direction. Without loss of generality, we can choose the $z$-direction in such a way, through an antisymmetric tensor, it is connected with the other two directions. Then, the tensor decomposition, given in Sec.~(\ref{subsec.tensormag}), is still valid. In this context, the antisymmetric tensor $\hat{F}^{\mu\nu}$ must be treated as an auxiliary element without any physical meaning.

Bearing in mind the above discussion, the vacuum result can be expressed in terms of a Ritus-like basis by projecting the effective vertex as in Eq.~(\ref{eq.coefa1}), giving
\begin{align}
    \label{eq.coefa1vac}
    a_{1,\text{vac.}}^{++}=&\mathcal{M}_\text{vac.}^{\mu\nu}\hat{L}_{1\mu}\hat{L}_{2\nu}=\dfrac{A_1}{|p_{1\perp}||p_{2\perp}|}\left(\left(p_1\cdot p_2\right)_\perp-\dfrac{p_2\hat{F}p_1}{p_1\cdot p_2}\right),\\
        \label{eq.coefa2vac}
        a_{2,\text{vac.}}^{++}=&\mathcal{M}_\text{vac.}^{\mu\nu}\hat{L}_{1\mu}^{*}\hat{L}_{2\nu}^{*}=\dfrac{A_1}{|p_{1\perp}||p_{2\perp}|}\left(\left(p_1\cdot p_2\right)_\parallel-\dfrac{p_2\hat{F}^{*}p_1}{p_1\cdot p_2}\right),\\
        \label{eq.coefa4vac}
    a_{4,\text{vac.}}^{+-}=&\mathcal{M}_\text{vac.}^{\mu\nu}\dfrac{1}{\sqrt{2}}\left(\hat{L}_{1\mu}\hat{L}_{2\nu}^{*}+\hat{L}_{1\mu}^{*}\hat{L}_{2\nu}\right)=-\dfrac{\sqrt{2}A_1}{|p_{1\perp}||p_{2\perp}|}\left(p_2\hat{F}p_1\right)\left(p_2\hat{F}^{*}p_1\right).
\end{align}
\noindent Once the above results are replaced in Eq.~(\ref{eq.sinB-PromEspin1}), the square amplitude depends only on $A_1$, as it is expected (see discussion bellow Eq.~\ref{eq.sinB-EstructuraVerticeFinal}).

\subsection{Strong magnetic field}
\label{subsec.apenIntegration.strong}

 \noindent In the strong magnetic field limit, discussed in Sec.~\ref{subsec.aproxStrong}, the diagram I  contribution to the amplitude, reads
\begin{equation}
	\begin{split}
		i\mathcal{M}_{E\hspace{1mm}\mu\nu}^{qB\hspace{0.5mm}(I)}(p_1,p_2)=&-ih g^2|qB|^{D_\perp/2}\dfrac{\left(-1\right)^{D_\perp}}{\pi^{D/2}4^{D_\perp-1}2^{D_\parallel-1}}\\
        &\times\int_{1/\Lambda^2}^{\infty} \ s^{2-D_\parallel/2} ds\int dv_1\ dv_2 \ dv_3 \ \delta(1-v_1-v_2-v_3) \ e^{-s \left(v_1v_3p_{1\parallel}^2+v_2v_3p_{2\parallel}^2+v_1v_2w^2_{\parallel}+m^2\right)}\\
		&\times\Bigg\{\textit{Tr}\left[\gamma_{\parallel\mu}\left(m+{v_3\slsh{p}_{1\parallel}+v_2\slsh{w}_{\parallel}}\right)\left(m-{v_1\slsh{w}_{\parallel}-v_3\slsh{p}_{2\parallel}}\right)\gamma_{\parallel\nu}\left(m-{v_1\slsh{p}_{1\parallel}+v_2\slsh{p}_{2\parallel}}\right)\Delta_+\right]\\
        &\hspace{86mm}-\frac{m}{2s}\Bigg((4-D_\parallel)\textit{Tr}\left[\gamma_{\parallel\mu}\gamma_{\parallel\nu}\Delta_+\right]\Bigg)\Bigg\}.
	\end{split}
	\label{eq.MF-VEdiagA4-StrongField2apen}
\end{equation}
\noindent By comparing the above expression with the zero magnetic field case, Eq.~(\ref{eq.MF-VEdiagAqB->0-1apen}), we observe a highly coincidence between the remaining integrals. Then, by using the incomplete Gamma function integral representation
\begin{equation}
	\int_{1/\Lambda^2}^{\infty}s^{n-\frac{D_\parallel}{2}}e^{-\varDelta_m^2s}ds=\left(\varDelta_m^2\right)^{\frac{D_\parallel}{2}-(n+1)}\Gamma\left(n+1-\dfrac{D_\parallel}{2},\frac{\varDelta_m^2}{\Lambda^2}\right),
	\label{eq.InteGammaIncomplete}
\end{equation}
\noindent in Eq.~(\ref{eq.MF-VEdiagA4-StrongField2apen}), the integration over the parameter $s$ can be easily performed, obtaining
\begin{equation}
	\begin{split}
		i\mathcal{M}_{E\hspace{1mm}\mu\nu}^{qB\hspace{0.5mm}(I)}&(p_1,p_2)=-i\dfrac{h g^2|qB|}{8\pi^2}\int_0^1 dv_1\int_0^{1-v_1} dv_2\\
		&\hspace{-9mm}\times\Bigg\{\Gamma\left(2,\frac{\Delta^2}{\Lambda^2}\right)\dfrac{\textit{Tr}\left[\gamma_{\parallel\mu}\left(m-(v_1-1)\slsh{p}_{1\parallel}+v_2\slsh{p}_{2\parallel}\right)\left(m-v_1\slsh{p}_{1\parallel}-(1-v_2)\slsh{p}_{2\parallel}\right)\gamma_{\parallel\nu}\left(m-v_1\slsh{p}_{1\parallel}+v_2\slsh{p}_{2\parallel}\right)\Delta_+\right]}{\left(m^2+v_1(1-v_1-v_2)p_{1\parallel}^2+v_2(1-v_1-v_2)p_{2\parallel}^2+v_1v_2w^2_{\parallel}\right)^2}\\
        &\hspace{55mm}-\Gamma\left(1,\frac{\Delta^2}{\Lambda^2}\right)\dfrac{m \ \textit{Tr}\left[\gamma_{\parallel\mu}\gamma_{\parallel\nu}\Delta_+\right]}{m^2+v_1(1-v_1-v_2)p_{1\parallel}^2+v_2(1-v_1-v_2)p_{2\parallel}^2+v_1v_2w^2_{\parallel}}\Bigg\},
	\end{split}
	\label{eq.MF-VEdiagA4-StrongField2apen3}
\end{equation}
\noindent where $\varDelta_m^2=m^2+v_1v_3p_{1\parallel}^2+v_2v_3p_{2\parallel}^2+v_1v_2w^2_{\parallel}$, the integration over the parameter $v_3$ has been additionally performed and the limit $D\longrightarrow4$ was straightforward taken because there are not divergent terms. As expected, in the strong magnetic field limit, there is a lineal dependence with the magnetic field, which can be related to the contributions of the LLL~\cite{Gusynin}.

By considering, the on-shell conditions and the hierarchy among the energy scales in the strong field limit $m^2, |p_{i\perp}|^2 \ \ll \ |qB|$, the incomplete Gamma functions can be approximated by its behavior around the study relevant point, which is 
\begin{equation}
  \dfrac{\Delta^2}{\Lambda^2}\sim -v_1 v_2\dfrac{m_\phi^2}{\Lambda^2}.
\end{equation}
\noindent At this point is clear that there is another energy scales comparison to be taken into account\footnote{Recall that the cut-off is an energy scale from which the field strength is the highest one.} $m_\phi^2\lessgtr|qB|$. Then, the remaining integrals over the parameters $v_1$ and $v_2$ can be performed in similar fashion as in the zero magnetic field case.

\subsection{Weak magnetic field approximation}
\label{subsec.apenIntegration.weak}

 \noindent As we discussed in Sec.~\ref{subsec.aproxWeak}, there are two different weak field approximations based on the kinematics of the external particles: low and high perpendicular momentum. The difference lies on the treatment of the exponential factor given in Eq.~(\ref{exptsai})
\begin{equation}
    \mathcal{X}(qB,s,v_j,p_{i\perp})=e^{-\frac{i}{qB}\frac{\mathrm{t_1}\mathrm{t_3}p_{1\perp}^2+\mathrm{t_2}\mathrm{t_3}p_{2\perp}^2+\mathrm{t_1}\mathrm{t_2}w_{\perp}^2+2\mathrm{t_1}\mathrm{t_2}\mathrm{t_3}p_2\hat{F}p_1}{\mathrm{t_1}\mathrm{t_2}\mathrm{t_3}-\mathrm{t_1}-\mathrm{t_2}-\mathrm{t_3}}},
\label{exptsaiapen}
\end{equation}
\noindent since it is a function of three energy scales and whose behavior depends on the hierarchy between $m$ and $p_{i\perp}$.

A power series expansion on $qBs$ can be performed on Eq.~(\ref{eq.MF-VEdiagA4})~\cite{TSAI.1,TSAI.2}, so, the general form for the weak field approximations reads
\begin{equation}
	\begin{split}
		i{\mathcal{M}_{{qB}}^{\mu\nu}}_{(I)}(p_1,p_2)\approx&ih g^2\left(-i\right)^{D_\perp+D/2+1}\dfrac{(-1)^{D_\perp/2}}{2^{D}\pi^{D/2}}\int dv_1\ dv_2 \ dv_3 \ \delta(1-v_1-v_2-v_3)\\
		&\times\int_0^\infty ds \ s^{2-D/2} \ e^{-ism^2}e^{is\left(v_1v_3p_{1}^2+v_2v_3p_{2}^2+v_1v_2w^2\right)}\mathcal{X}_{\text{low/high}}(qB,s,v_j,p_{i\perp}) \ f(qBs,v_i,p_{i\perp}),
	\end{split}
	\label{eq.MF-VEdiagAweakfield}
\end{equation}	
\noindent where the functions $\mathcal{X}_{\text{low/high}}(qB,s,v_j,p_{i\perp})$ correspond to the weak field at low or high perpendicular momentum approximation, respectively and $f(qBs,v_i,p_{i\perp})$ is a polynomial function on $qBs$, with the variables used in Sec.~\ref{sec.aproximations}.

In the weak field with low perpendicular momentum approximation, the exponential factor in Eq.~(\ref{exptsaiapen}) can be expanded in a Taylor series, so the leading terms, up to $\mathcal{O}\left((qB)^2\right)$, are the following
\begin{equation}
    \begin{split}
        \mathcal{X}_{\text{low}}(qB,s,v_j,p_{i\perp})=1+&i2qBs^2v_1v_2v_3 \ p_2\hat{F}p_1-2(qB)^2s^4v_1^2v_2^2v_3^2 \ p_2\hat{F}p_1\\
        &+i\frac{(qB)^2s^3}{3}\left(-v_2^2+v_1v_2+v_1v_3+v_2v_3\right)v_1v_3p_{1\perp}^2\\
        &\hspace{2mm}+i\frac{(qB)^2s^3}{3}\left(-v_1^2+v_1v_2+v_1v_3+v_2v_3\right)v_2v_3p_{2\perp}^2\\
        &\hspace{4mm}+i\frac{(qB)^2s^3}{3}\left(-v_3^2+v_1v_2+v_1v_3+v_2v_3\right)v_1v_2w_{\perp}^2.
        \label{eq.xlow}
    \end{split}
\end{equation}

Next, in this approximation, the integration over the parameter $s$ in Eq.~(\ref{eq.MF-VEdiagAweakfield}) can be easily performed, term by term, by using the integral representation of the Gamma function, Eq.~(\ref{eq.sinB-InteGamma1}). Note that the terms with non-vanishing power of $qB$ contain extra factors of $s$ in the numerator, which improves its behavior in the $s\longrightarrow0$ neighborhood (UV region).

By considering the on-shell conditions, the remaining integrals over $v_1$ and $v_2$ can be performed analytically in similar fashion as Eqs.~(\ref{eq.A1vacio})-(\ref{eq.A2vacio}).

In the weak field approximation with high perpendicular momentum, we cannot a perform a Taylor series expansion in the exponential term in Eq.~(\ref{exptsaiapen}), since there are factors of $p_{i\perp}$ that make the exponential argument larger than one. However, a power expansion on its argument is valid, giving
\begin{equation}
    \begin{split}
        \mathcal{X}_{\text{high}}(qB,s,v_j,p_{i\perp})=e^{i2qBs^2v_1v_2v_3 \ p_2\hat{F}p_1}&e^{i\frac{(qB)^2s^3}{3}\left(-v_2^2+v_1v_2+v_1v_3+v_2v_3\right)v_1v_3p_{1\perp}^2}\\
    &\times e^{i\frac{(qB)^2s^3}{3}\left(-v_1^2+v_1v_2+v_1v_3+v_2v_3\right)v_2v_3p_{2\perp}^2}\\
    &\hspace{4mm}\times e^{i\frac{(qB)^2s^3}{3}\left(-v_3^2+v_1v_2+v_1v_3+v_2v_3\right)v_1v_2w_{\perp}^2}.
    \end{split}
     \label{eq.xhigh}
\end{equation}

In this physical regime, the $qB$ dependence is not fully contained in the polynomial function and the integration can not be performed, as in the previous case, by using an integral representation of the Gamma functions, due to extra exponential factors with higher powers on $s$. The method to perform this kind of integrals is a work in progress and will be report elsewhere.

\acknowledgements

Support for this work has been received in part from DGAPA-UNAM under grant number PAPIIT-IN108123.


\bibliographystyle{unsrt}
\bibliography{BIBLIO2}

\end{document}